\documentclass{aa}
\usepackage{graphicx}
\usepackage{etex}
\usepackage[T1]{fontenc}
\usepackage[latin1]{inputenc}
\usepackage{units}

\usepackage{graphics}
\def\bdm{\begin{displaymath}}
\def\edm{\end{displaymath}}

\newcommand{\nid}{\noindent}
\newcommand{\pd}[1]{\partial #1}
\newcommand{\td}[1]{\, \mathrm{d} #1 \,}
\newcommand{\intl}{\int\limits}

\newcommand{\HSF}[1]{\; \mathrm{H}\left[ #1 \right] }
\newcommand{\DELF}[1]{\; \delta\left( #1 \right) }
\def\be{\begin{eqnarray}}
\def\ee{\end{eqnarray}}
\newcommand{\eqd}{\,\, .}
\newcommand{\eqc}{\,\, ,}
\newcommand{\ggg}{g_2^{-1}}

\newcommand{\xc}{\tilde{x}}
\newcommand{\yc}{\tilde{y}}
\newcommand{\ys}{\tilde{y}^{\ast}}
\newcommand{\cc}{\tilde{c}}
\newcommand{\Fa}{F_{\alpha}}
\newcommand{\Fc}{F_{COM}}
\newcommand{\Psis}{\Psi^{\ast}}
\newcommand{\Fs}[1]{F_c^{\ast #1}}
\newcommand{\Us}{\tilde{U}}
\begin{document}
\title{Blazar synchrotron emission of instantaneously power-law injected electrons under 
linear synchrotron, non-linear SSC, and combined synchrotron-SSC cooling}
\author{M. Zacharias \& R. Schlickeiser }
\institute{ Institut für Theoretische Physik, Lehrstuhl IV:
Weltraum- und Astrophysik, Ruhr-Universität Bochum, 44780 Bochum, Germany}
\date{Received ?; accepted ? }
\offprints{M. Zacharias; mz@tp4.rub.de}
%
\authorrunning{Zacharias \& Schlickeiser}
\titlerunning{Nonlinear SSC electron cooling}
\abstract
{The broadband SEDs of blazars show two distinct components which in leptonic models are associated with synchrotron 
and SSC emission of highly relativistic electrons. In some sources the SSC component dominates the synchrotron peak by one 
or more orders of magnitude implying that the electrons mainly cool by inverse Compton collisions with their self-made 
synchrotron photons. Therefore, the linear synchrotron loss of electrons, which is normally invoked in emission models, has to 
be replaced by a nonlinear loss rate depending on an energy integral of the electron distribution. This modified 
electron cooling changes significantly the emerging radiation spectra.}   
%
{It is the purpose of this work to apply this new cooling scenario to relativistic power-law distributed electrons, which are 
injected instantaneously into the jet.}   
{We will first solve the differential equation of the volume-averaged differential number density of the electrons, and then 
discuss their temporal evolution. 
Since any non-linear cooling will turn into linear cooling after some time, we also calculated the electron number density for a combined cooling scenario consisting of both the linear and non-linear cooling. \\
For all cases, we will also calculate analytically the emerging optically thin synchrotron fluence spectrum which will be compared to a numerical solution.}   
{For small normalized frequencies $f<1$ the fluence spectra show constant spectral indices. We find for linear 
cooling $\alpha_{SYN}=1/2$, and for non-linear cooling $\alpha_{SSC}=3/2$. In the combined cooling scenario we obtain for the 
small injection parameter $\beta_1=1/2$, and for the large injection parameter $\beta_2=3/2$, which 
becomes $\beta_1=1/2$ for very small frequencies, again. These identical behaviors, as compared to the existing calculations for
monoenergetically injected electrons, prove that the spectral behavior of the total synchrotron fluence is independent from the 
functional form of the energy injection spectrum.}   
{}   
\maketitle

\section{Introduction}

Amongst the brightest sources visible throughout the observable universe are blazars, the most extreme class of 
Active Galactic Nuclei (AGN). 
AGN are galaxies whose central super massive black hole (SMBH) is fed by tremendous amounts of gas, which accumulate in an 
accretion disk surrounding the SMBH. Magnetic fields trapped in the plasma are swirled around by the rotating disk 
forming two narrow channels, commonly known as jets, along the rotational axis of the SMBH. Although the launching process 
in all its details is not yet fully understood (Spruit (2010)), it is well established that it is an important process 
to remove angular momentum from the disk. The angular momentum is carried away by particles moving at highly relativistic 
speeds through the jets. 

In most jet models electrons (negatrons and positrons) form the particle content of the jet, however heavier hadronic components 
might also be present. The charged particles are subject to several radiation processes, and due to the relativistic 
motions the emerging radiation is effectively beamed in the forward direction. Thus, if an AGN is observed straight down the jet, 
it is extremely bright and is called a blazar. 

The observed photon energy spectrum of a blazar is dominated by two broad components of nonthermal radiation. In leptonic 
models (for a review see Böttcher (2007)) the low-energetic component is attributed to synchrotron radiation of 
highly relativistic electrons, while the high-energetic component is due to inverse Compton collisions of the electrons with 
ambient radiation fields. Such radiation fields could be of external nature (e. g. Dermer \& Schlickeiser (1993)), like the cosmic 
microwave background, radiation directly from the accretion disk, or photons from the accretion disk scattered in the infrared 
torus surrounding the disk or in the Broad- or Narrow-Line-Regions circling the SMBH. 
However, since the electrons will in any case produce synchrotron radiation due to their interaction with the magnetic field 
of the jet, it is unavoidable that they Compton-scatter their self-made synchrotron photons up to $\gamma$-ray energies. This is 
called the synchrotron self-Compton (SSC) mechanism. 

Another very important topic concerning blazars is the short time variaility of their emission at practically all photon energies. 
Blazars have shown variabilities on the order of minutes (Aharonian et al. (2007), Ghisellini et al. (2009a)), and such 
rapid variabilities have to be explained by radiation and emission models. In this respect, 
some analytical work has been done recently regarding the possibility of very rapid nonlinear radiation processes, which could be 
at work in leptonic radiation models of jets. Schlickeiser \& Lerche (2007,2008) and Zacharias \& Schlickeiser (2009) discussed a 
nonlinear synchrotron model which assumed equipartition between the magnetic field and the electron energy density. 
Schlickeiser (2009, hereafter referred to as RS) investigated as additional nonlinear cooling process the synchrotron self-Compton 
radiation process, which is described above. 

In many Blazars, like PKS 0048-071, PKS 0202-17, PKS 0528+134, 3C 454.3 (Ghisellini, Tavecchio \& Ghirlanda (2009)), 
S5 0014+813 (Ghisellini et al. (2009b)), and others from the Fermi blazar survey (Abdo et al. 2010), the Compton peak dominates by at least one order of magnitude over the synchrotron peak, which is even more pronounced when the $\gamma$-ray absorbtion of the extragalactic background light (EBL) is factored in (e.g. Venters, Pavlidou \& Reyes (2009)). \\
Assuming that the Compton peak is mainly a result of SSC radiation one finds for the ratio of the luminosities of the peaks (Schlickeiser, Böttcher \& Menzler (2010), hereafter referred to as SBM)
\be
\frac{L_{SSC}^{*}}{L_{SYN}^{*}}=\frac{\int\td{V}\int_1^{\infty}\td{\gamma}n(\gamma)|\dot{\gamma}_{SSC}|}{\int\td{V}\int_1^{\infty}\td{\gamma}n(\gamma)|\dot{\gamma}_{SYN}|} \eqd
\ee
Since SSC cooling relies on the synchrotron photons of the same electrons, the electron density $n(\gamma)$ is the same in 
both cases. Similarly, the Doppler boosting factors are also identical (Dermer \& Schlickeiser (2002)), which means that 
the ratio of the luminosities directly reflects the ratio of the cooling factors $|\dot{\gamma}_{i}|$ (with $i\in\{SSC,SYN\}$). 
Hence, a dominance of the Compton peak over the synchrotron peak implies that the electrons mainly cool by inverse Compton collisions 
with their self-made photons. In this case, the linear synchrotron cooling has to be replaced by another cooling process 
dealing with SSC radiation, which has been considered by RS. 

It is important to notice that the nonlinear SSC cooling only operates in the Thomson regime and does not deal with higher order 
SSC collisions, which are possible. Taking higher oder interactions and/or the Klein-Nishina (KN) regime into account, would 
lead to a much more complicated approach. If KN effects are important in the beginning, our approach would be at least a good 
approximation for later times, since the electrons cool significantly over time and will sooner or later reach the 
Thomson scattering regime. However, blazar spectral energy distributions (SEDs) normally do not show a third broad component, 
which would be an indicator of higher order scattering or scattering in the KN domain. This is a hint for the validity of the 
approach. We notice that this restriction to the Thomson regime requires according to RS that the maximum Lorentz factor 
is not larger than $\gamma_{max}\approx 1.94\cdot 10^4 b^{-1/3}$ ($b$ is the normalized magnetic field strength in units of Gauss). 

RS applied the nonlinear SSC model to the simple but illustrative case of mono-energetic electrons that are instantaneously injected into the jet. It is the purpose of this work to apply this new model to a more realistic scenario. We assume an instantaneous injection of power-law distributed electrons of the form $Q(\gamma,t)\propto \gamma^{-s}\DELF{t}$. This describes a single flare event, in which one population of electrons is injected into the jet and causes a sudden outburst of radiation. We should also note that we treat only the optically thin case here. \\
We will begin with the calculation of the electron number densities in the linear and nonlinear case in section \ref{sec:end}. We will also discuss the temporal evolution of the respective electron densities and will outline their behavior by evaluating the upper and lower cut-offs of the spectra. We will show that any non-linear cooling will ultimately become linear after some time, which requires a treatment with both cooling scenarios in one equation. This has been carried out for mono-energetic electrons by SBM, and we will extend their research here, as well, in section \ref{sec:ccend}. \\
Sections \ref{sec:lsf}, \ref{sec:sfs} and \ref{sec:sfcom} are devoted to the calculation of the total linear and nonlinear synchrotron  fluence and the fluence of the combined cooling scenario, respectively. We will discuss our results in section \ref{sec:dac}.


\section{Linear and non-linear electron number densities} \label{sec:end}

In this section we calculate the distribution function of the cooled electrons for linear synchrotron and nonlinear SSC cooling, respectively, and discuss their temporal evolution. \\
The electron number density $n(\gamma,t)$ is governed by the competition between time-varying energy losses and the injection of relativistic electrons into the jet. This competition is described by the partial differential equation (Kardashev (1962))
\be
\frac{\pd{n(\gamma,t)}}{\pd{t}}-\frac{\pd}{\pd{\gamma}}\left[ |\dot{\gamma}|n(\gamma,t) \right] = Q(\gamma,t) \label{eldispde} \eqd
\ee
$Q(\gamma,t)$ is the injection rate and we assume an instantaneous injection of power-law distributed electrons in the form
\be
Q(\gamma,t)=q_0\gamma^{-s}\HSF{\gamma-\gamma_1}\HSF{\gamma_2-\gamma}\DELF{t} \label{injecteldis} \eqc
\ee
where $s$ is the spectral index, $\gamma_1$ and $\gamma_2$ are the lower and upper cut-offs of the electron spectrum, respectively, $\HSF{x}$ denotes Heaviside's step function, and $\DELF{x}$ is Dirac's $\delta$-distribution.


\subsection{Linear cooling} \label{ssec:lc}

\begin{figure}[ht]
	\centering
		\includegraphics[width=0.50\textwidth]{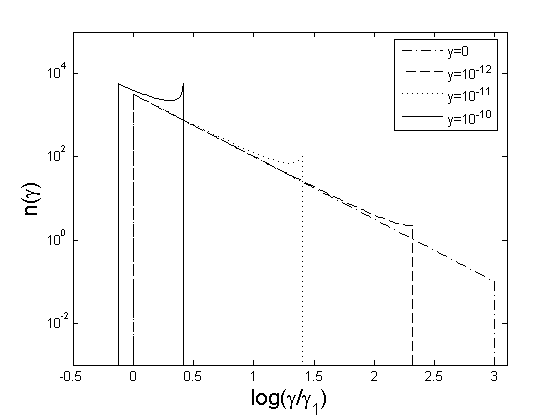}
	\caption{Linear electron distribution for $s=1.5$ at four different times $y$ for the initial values: $\gamma_1=10^1$, and $\gamma_2=10^4$.}
	\label{fig:ElDis15syn}
\end{figure}

\begin{figure}[ht]
	\centering
		\includegraphics[width=0.50\textwidth]{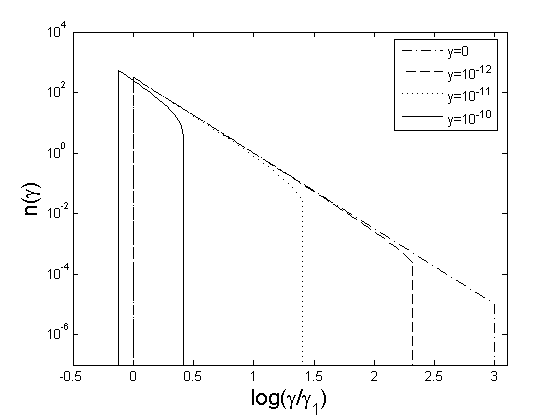}
	\caption{Linear electron distribution for $s=2.5$ at four different times $y$ for the initial values: $\gamma_1=10^1$, and $\gamma_2=10^4$.}
	\label{fig:ElDis25syn}
\end{figure}

\begin{figure}[ht]
	\centering
		\includegraphics[width=0.50\textwidth]{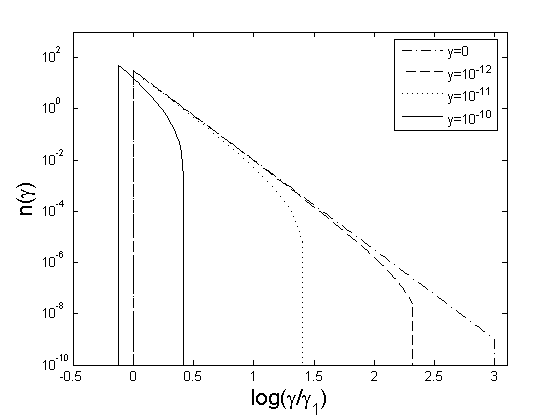}
	\caption{Linear electron distribution for $s=3.5$ at four different times $y$ for the initial values: $\gamma_1=10^1$, and $\gamma_2=10^4$.}
	\label{fig:ElDis35syn}
\end{figure}

The linear cooling is described by the linear energy loss term
\be
|\dot{\gamma}|_{SYN}=\frac{4c\sigma_T}{3mc^2}U_B\gamma^2\equiv D_0\gamma^2 \label{syncool} \eqc
\ee
where $U_B=B^2/8\pi$ is the magnetic energy density of a magnetic field $B=b\,\unit{G}$. Schlickeiser \& Lerche (2008) obtained the solution for the linear cooling:
\be
n_{SYN}(\gamma,t)=q_0\gamma^{-s}(1-D_0\gamma t)^{s-2}\HSF{\gamma-\frac{\gamma_1}{1+D_0\gamma_1 t}} \nonumber \\ 
\times \HSF{\frac{\gamma_2}{1+D_0\gamma_2 t}-\gamma} \label{lineldis} \eqd
\ee
The temporal evolution (with $y=A_0 t$, and $A_0$ defined in the next subsection) of this solution is shown in Figs. \ref{fig:ElDis15syn} - \ref{fig:ElDis35syn} for three different cases of $s$.
 \\
 \\
 \\
As one can see, the electron distributions piles up at the high end of the spectrum for spectral indices $s<2$, as it was already discovered by Kardashev (1962). This does not happen for larger $s$, and in fact these cases show the exact opposite behavior with a drop at the high-energy end. It is also obvious that $s$ only determines the steepness of the spectrum, and not the cooling time scales.


\subsection{Nonlinear cooling} \label{ssec:nlc}

According to RS the nonlinear SSC cooling rate is defined by
\be
|\dot{\gamma}|_{SSC}=A_0\gamma^2\intl_{0}^{\infty}\td{\gamma}\gamma^2n(\gamma,t) \label{ssccool} \eqc
\ee
with the constant $A_0=3c_1\sigma_TP_0RE_k^2/mc^2$. Here $\sigma_T$ is the Thomson cross-section of electron scattering, $R$ is the radius of the spherical source, $P_0=\alpha_f/2\sqrt{3}\hbar=3.2\cdot 10^{12}\,\unit{eV^{-1}s^{-1}}$, $E_k=1.16\cdot 10^{-8}b\,\unit{eV}$, $c_1\approx 0.684$, and $mc^2$ is the rest-mass energy of an electron. \\
The integral over the electron number density $n(\gamma,t)$ basically introduces a time dependence of the cooling rate. This is reasonable, since cooler electrons have less energy they can transfer to the photons, which implies that the electrons do not cool effectively any more. \\
The solution to the differential equation (\ref{eldispde}) is derived in appendix \ref{sec:append}, and becomes
\be
n_{SSC}(\gamma,T)=q_0\gamma^{-s}\left( 1-\gamma T \right)^{s-2}\HSF{\frac{\gamma_2}{1+\gamma_2 T} -\gamma} \nonumber \\
\times \HSF{\gamma- \frac{\gamma_1}{1+\gamma_1 T}}  \label{nleldis} \eqd
\ee
The implicit time variable $T$ is defined by the nonlinear differential equation
\be
\frac{\td{T}}{\td{y}}=\intl_{0}^{\infty}\td{\gamma}\gamma^{2}n_{SSC}(\gamma,T) \label{impltimevariab} \eqc
\ee
where $y=A_0t$. As we will see later the derivative of $T$ with respect to $y$ is as important as the complete time variable $T$ itself in order to calculate the synchrotron spectra (cf. section \ref{sec:sfs}). The calculation of the former is presented in appendix \ref{sec:appu}, and becomes for $s>3$
\be
U(x) = \left\{ \begin{array}{ll} 
U_0 \left[ 1-g_2^{3-s} \right] & \eqc\, 0\leq x\leq g_2^{-1} \\ 
U_0 \left[ 1-\frac{2x^{s-3}}{s-1}-\frac{s-3}{s-1}\frac{g_2^{1-s}}{x^2} \right] & \eqc\, g_2^{-1}\leq x\leq 1 \\
U_0 \frac{s-3}{(s-1)x^2}\left[ 1-g_2^{1-s} \right] & \eqc\, x\geq 1  
\end{array} \right. \label{impltimesol1} \eqc
\ee
and for $1<s<3$
\be
U(x) = \left\{ \begin{array}{ll} 
U_1 \left[ g_2^{3-s}-1 \right] & \eqc\, 0\leq x\leq g_2^{-1} \\ 
U_1 \left[ \frac{2x^{s-3}}{s-1}-1-\frac{s-3}{s-1}\frac{g_2^{1-s}}{x^2} \right] & \eqc\, g_2^{-1}\leq x\leq 1 \\
U_1 \frac{3-s}{(s-1)x^2}\left[ 1-g_2^{1-s} \right] & \eqc\, x\geq 1  
\end{array} \right. \label{impltimesol2} \eqc
\ee
with the new time variable $x=\gamma_1 T$, the ratio between the initial cut-offs $g_2=\gamma_2/\gamma_1$, and the constants $U_0=\frac{q_0\gamma_1^{3-s}}{s-3}$, and $U_1=\frac{q_0\gamma_1^{3-s}}{3-s}$.  


\subsubsection{Temporal evolution of the nonlinear electron number density} \label{ssec:teend}

If one is interested in the actual temporal evolution of the electron distribution, one has to perform the integration of $U(x)$ with respect to the time-variable, yielding for $s>3$
\be
x(y) = \nonumber \\ 
\left\{ \begin{array}{ll} 
\gamma_1 U_0 \left[ 1-g_2^{3-s} \right] y & \eqc\, 0\leq y\leq y_1 \\ 
\gamma_1 U_0 \left[ y-c_2 \right] & \eqc\, y_1\leq y\leq y_2 \\
\left[ 3\gamma_1 U_0 \frac{s-3}{s-1} (1-g_2^{1-s})(y-c_3) \right]^{1/3} & \eqc\, y\geq y_2  
\end{array} \right. \label{timevabsol1} \eqc
\ee
and for $1<s<3$
\be
x(y) = \nonumber \\
\left\{ \begin{array}{ll} 
\gamma_1 U_1 \left[ g_2^{3-s}-1 \right]y & \eqc\, 0\leq y\leq y_3 \\ 
\left[ 2\gamma_1 U_1 \frac{4-s}{s-1}(y-c_5) \right]^{\frac{1}{4-s}} & \eqc\, y_3\leq y\leq y_4 \\
\left[ 3\gamma_1 U_1 \frac{3-s}{s-1}(1-g_2^{1-s})(y-c_6) \right]^{1/3} & \eqc\, x\geq y_4  
\end{array} \right. \label{timevabsol2} \eqd
\ee
The integration and the approximations used for it can be found in appendix \ref{sec:appx}. The time limits $y_i$ and the constants $c_i$ are chosen in such a way that the solution $x(y)$ is continuous for all $y$. The values are also listed in appendix \ref{sec:appx}. 
\begin{figure}[ht]
	\centering
		\includegraphics[width=0.50\textwidth]{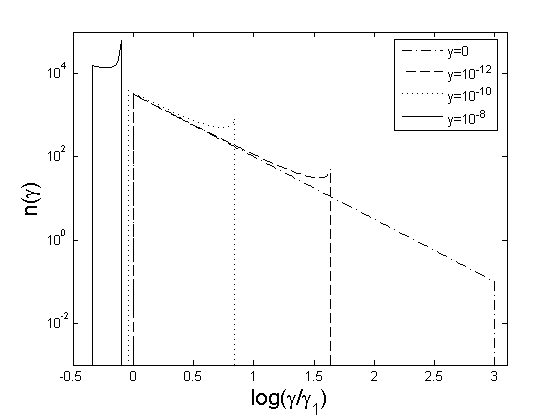}
	\caption{Nonlinear electron distribution for $s=1.5$ at four different times $y$ for the initial values: $\gamma_1=10^1$, and $\gamma_2=10^4$.}
	\label{fig:ElDis15}
\end{figure} \\
\begin{figure}[ht]
	\centering
		\includegraphics[width=0.50\textwidth]{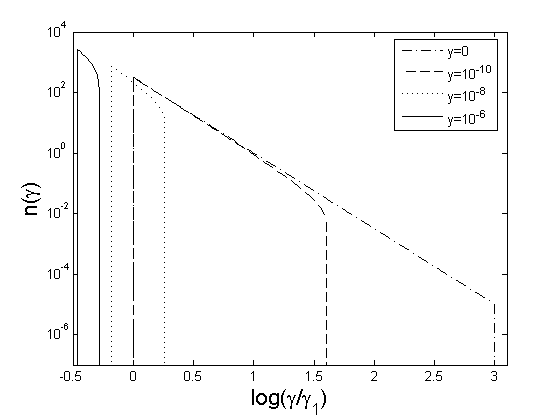}
	\caption{Nonlinear electron distribution for $s=2.5$ at four different times $y$ for the initial values: $\gamma_1=10^1$, and $\gamma_2=10^4$.}
	\label{fig:ElDis25}
\end{figure} \\
\begin{figure}[ht]
	\centering
		\includegraphics[width=0.50\textwidth]{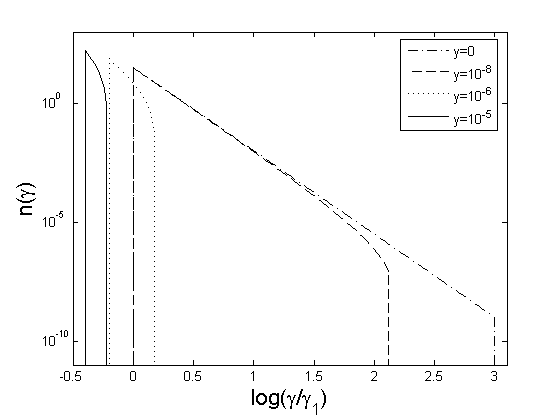}
	\caption{Nonlinear electron distribution for $s=3.5$ at four different times $y$ for the initial values: $\gamma_1=10^1$, and $\gamma_2=10^4$.}
	\label{fig:ElDis35}
\end{figure} \\
In Figs. \ref{fig:ElDis15} - \ref{fig:ElDis35} we plotted the time evolution of the nonlinear electron spectrum for $s=1.5$, $s=2.5$, and $s=3.5$, respectively. \\
As one can see, the spectrum cools significantly faster for harder spectra (smaller spectral index). It is also quite obvious that the spectrum contracts rapidly, which means that for late times the spectrum is identical to a $\delta$-function (cf. next subsection). One should also notice the increasing pile-up for $s<2$ at the upper limit for later times, which becomes a drop for larger $s$, similarly to the linear case.


\subsection{Temporal evolution of the cut-offs} \label{ssec:teco}

\begin{figure}[ht]
	\centering
		\includegraphics[width=0.50\textwidth]{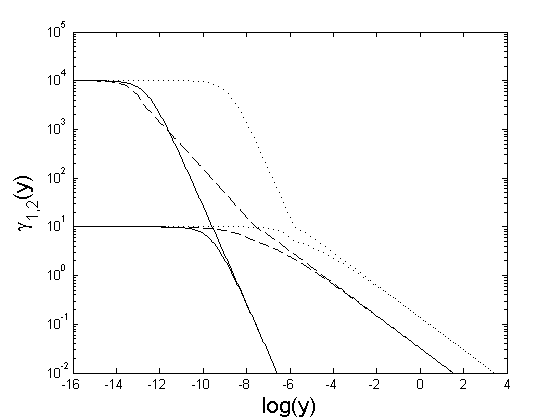}
	\caption{Time dependence of the upper ($\gamma_2$, upper curves) and lower ($\gamma_1$, lower curves) cut-offs of the electron spectra, respectively, for the standard values $\gamma_1=10^1$, $\gamma_2=10^4$, $R_{15}=1$, and $q_5=1$. \textit{Full curve:} Linear electron cooling for $s=2$. \textit{Dashed curve:} Nonlinear electron cooling for $s=2$. \textit{Dotted curve:} Nonlinear electron cooling for $s=3.5$.}
	\label{fig:heavi2035syn}
\end{figure}

The qualitative behavior outlined in the last subsections of the electron densities is confirmed by Fig. \ref{fig:heavi2035syn} showing the time dependence of the upper and lower cut-offs defined by the Heaviside functions of the respective electron densities.  
 \\
Since the nonlinear cooling depends critically on $s$, we plotted it for the two cases $s=2$, and $s=3.5$; for the linear case we used $s=2$, but as we have seen before this choice is not important. For early times the cut-offs of the nonlinear scenario remain practically unchanged. However, the cooling of the high-energetic electrons sets in about three orders of magnitude earlier than the cooling of the low-energetic ones resulting in the quenching of the spectrum, which leads to an electron spectrum at later times that basically looks like a $\delta$-function. This is expected from the denominator of the Heaviside functions in the electron number density, because of the factor $g_2=10^3$ in these examples. It is also obvious that at these later times the nonlinear electron distributions cool with a $y^{-1/3}$-dependence. In the intermediate time range the cooling of the softer spectrum (larger spectral index) is faster than for the harder case. \\
Concerning the linear cooling case, we find that once the cooling has begun, the electrons cool with a $y^{-1}$-dependence as expected. Similarly to the nonlinear case, the cooling of the high-energetic electrons begins a factor of about $\ggg$ earlier than for the low-energetic ones, resulting again in a $\delta$-function appearance of the electron spectrum for later times. 
\begin{figure}[ht]
	\centering
		\includegraphics[width=0.50\textwidth]{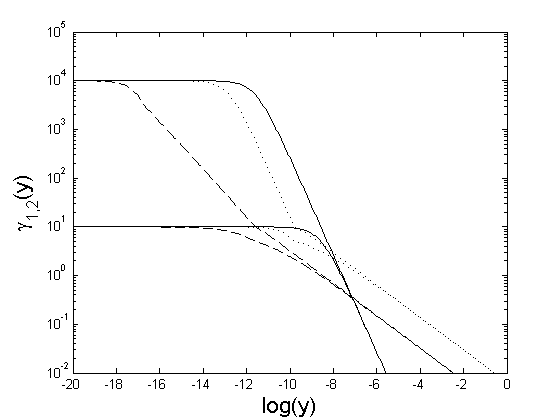}
	\caption{The same as in Fig. \ref{fig:heavi2035syn},  but with $R_{15}=10$, and $q_5=10^4$. Note the different values on the horizontal axis.}
	\label{fig:heavi2035syn2}
\end{figure} \\
In Fig. \ref{fig:heavi2035syn} the linear cooling operates much faster than the nonlinear cooling. This depends critically on the parameters we have chosen for the plot: $R=10^{15}R_{15}\,\unit{cm}$, and $q_0=10^{5}q_5\,\unit{cm^{-3}}$. Especially for higher values of these parameters (cf. Fig. \ref{fig:heavi2035syn2}) the linear cooling begins at much later times for larger $R$, or the nonlinear cooling begins much earlier for larger $q_0$ (the magnetic field strength does not change the overall result, since both cooling scenarios depend on $B^2$; cf. the definitions of $D_0$ and $A_0$). We note, however, that at one point in time the linear cooling will always dominate over the nonlinear cooling, since it cools faster for later times than the nonlinear case. \\
This behavior indicates that linear and nonlinear cooling should not be treated separately, as we did in this section. This fact has been treated by SBM for the illustrative case of mono-energetic electrons (at all times). We will calculate the number density of the combined linear and non-linear cooled (hereafter referred to as "combined" cooling) electrons in the next section.


\section{Combined cooled electron number density} \label{sec:ccend}

The combined cooling is described by the energy loss term
\be
|\dot{\gamma}|_{COM}=|\dot{\gamma}|_{SYN}+|\dot{\gamma}|_{SSC} \nonumber \\
=A_0\gamma^2\left( K_0+\intl_0^{\infty}\td{\gamma}\gamma^2 n_{COM}(\gamma,t) \right) \label{comcool} \eqc
\ee
with $K_0=D_0/A_0$. \\
Inserting this into Eq. (\ref{eldispde}) yields the solution
\be
n_{COM}(\gamma,\tau)=q_0\gamma^{-s}\left( 1-\gamma \tau \right)^{s-2}\HSF{\frac{\gamma_2}{1+\gamma_2 \tau} -\gamma} \nonumber \\
\times \HSF{\gamma- \frac{\gamma_1}{1+\gamma_1 \tau}}  \label{comeldis} \eqd
\ee
The similarity of Eq. (\ref{comeldis}) with Eq. (\ref{nleldis}) is not surprising, since the way to derive the solution is the same in both ways. The important difference is that the implicit time variable $T$ has changed to the new definition $\tau$:
\be
\frac{\td{\tau}}{\td{y}}=K_0+\intl_{0}^{\infty}\td{\gamma}\gamma^{2}n_{COM}(\gamma,\tau) \label{comimpltimevariab} \eqd
\ee
In appendix \ref{sec:appxc} we derive the complete solution of the implicit time variable $\xc=\gamma_1\tau$, which becomes for $s>3$
\be
\xc(y,\alpha_0\ll 1)=\nonumber \\ 
\left\{ \begin{array}{ll} 
\gamma_1 (K_0+U_0) y & \eqc\, 0\leq y\leq \yc_2 \\ 
\gamma_1 K_0 (y-\cc_4) & \eqc\, y\geq \yc_2 
\end{array} \right. \label{comtimevabsol1} \eqc
\ee
\be
\xc(y,\alpha_0\gg 1)=\nonumber \\ 
\left\{ \begin{array}{ll} 
\gamma_1 (K_0+U_0) y & \eqc\, 0\leq y\leq \yc_2 \\ 
\left[ 3\alpha_0^2\gamma_1 K_0 (y-\cc_5) \right]^{1/3} & \eqc\, \yc_2 \leq y\leq \yc_3 \\
\gamma_1 K_0 (y-\cc_6) & \eqc\, y\geq\yc_3 
\end{array} \right. \label{comtimevabsol2} \eqc
\ee
and for $1<s<3$
\be
\xc(y,\alpha_0\ll 1)=\nonumber \\ 
\left\{ \begin{array}{ll} 
\gamma_1 (K_0+U_1 g_2^{3-s}) y & \eqc\, 0\leq y\leq \ys_1 \\ 
\frac{2\gamma_1 U_1\beta}{(s-1)\alpha_0^2}(y-d_2) & \eqc\, \ys_1\leq y\leq\ys_2 \\
\gamma_1 K_0 (y-d5) & \eqc\, y\geq\ys_2
\end{array} \right. \label{comtimevabsol3} \eqc
\ee
\be
\xc(y,\alpha_0\gg 1)=\nonumber \\ 
\left\{ \begin{array}{ll} 
\gamma_1 (K_0+U_1 g_2^{3-s}) y & \eqc\, 0\leq y\leq \ys_1 \\
\left[ \frac{2\gamma_1 U_1 (4-s)}{(s-1)}(y-d_3) \right]^{\frac{1}{4-s}} & \eqc\, \ys_1\leq y\leq \ys_3 \\
\left[ 3\alpha_0^2\gamma_1 K_0 (y-d_6) \right]^{1/3} & \eqc\, \ys_3 \leq y\leq \ys_4 \\
\gamma_1 K_0 (y-d_7) & \eqc\, y\geq\ys_4 
\end{array} \right. \label{comtimevabsol4} \eqd
\ee
The approximations and the values of the constants $\yc_i$, $\ys_i$, $\cc_i$, and $d_i$ can be found in appendix \ref{sec:appxc}, as well. They are chosen in such a way that $\xc$ is continuous for all times. There we also introduce $\beta=(3-s)(1-g_2^{1-s})/2$ and the important injection parameter
\be
\alpha_0 = \sqrt{\frac{q_0\gamma_1^{3-s}(1-g_2^{1-s})}{(s-1)K_0}}\simeq \sqrt{\frac{q_0\gamma_1^{3-s}}{(s-1)K_0}} \label{injparam} \eqc
\ee
for extended power law injection spectra. The power law injection parameter $\alpha_0$ depends on the particle number density $q_0$, the initial lower cut-off $\gamma_1$, the electron spectral index $s$, the constant $K_0$, and only weakly on the upper cut-off, since $g_2^{1-s}\ll 1$. 


\subsection{Power law injection parameter} \label{ssec:injpar}

For standard blazar parameters $q_0=10^5q_5\,\unit{cm^{-3}}$ and $R=10^{15}R_{15}\,\unit{cm}$ we use from SBM $(K_0/q_0)^{1/2}=106(R_{15}q_5)^{-1/2}$ and introduce the total number of instantaneously injected electrons 
\be
N=10^{50}N_{50}={4\pi \over 3}R^3q_0\int _{\gamma _1}^{g_2\gamma _1}d\gamma \gamma ^{-s} \nonumber \\
\simeq 4.2\cdot 10^{50}q_5R_{15}^3{\gamma _1^{1-s}\over s-1} \eqc \label{elecnum}
\ee
yielding $q_5=(s-1)N_{50}/(4.2R_{15}^3\gamma _1^{1-s})$, so that the power law injection parameter becomes 
\be
\alpha_0 ={\gamma _1\over \gamma _B}=4.6\cdot 10^{-3}{\gamma _1N_{50}^{1/2}\over R_{15}}
\label{ip1}
\ee
in terms of the characteristic Lorentz factor introduced by SBM 
\be
\gamma _B={217R_{15}\over N_{50}^{1/2}}
\label{ip2}
\ee
The injection parameter is determined by the total number of injected electrons and the size of the source as indicated, but independent of the magnetic field strength. If the lower cutoff of the injected power law $\gamma _1$ is higher (smaller) than $\gamma _B$, the injection parameter $\alpha_0 $ will be larger (smaller) than unity. For a compact source with a large number of injected electrons, the injection parameter is much larger than unity. \\
Basically the injection parameter is the fundamental ordering parameter that defines the ratio between the magnitudes of the synchrotron and 
the Compton peak. An injection parameter smaller than unity corresponds to a higher synchrotron flux compared to the Compton flux leading to the linear synchrotron cooling of the electrons, as one can see in Eqs. (\ref{comtimevabsol1}) and (\ref{comtimevabsol3}). Similarly, an injection parameter being larger than unity leads to the non-linear SSC cooling shown in Eqs. (\ref{comtimevabsol2}) and (\ref{comtimevabsol4}), since the flux of the Compton peak is higher than of the synchrotron peak, which was the condition for the non-linear cooling. One should note, however, that our remarks from the previous section hold: Any non-linear cooling will sooner or later turn into linear cooling. \\
It is also noteworthy that $\alpha_0$ is independent of the magnetic field strength. This results from the same dependence 
of both, $A_0$ and $D_0$, on $b^2$. 


\subsection{Temporal evolution of the combined electron number density} \label{ssec:comtecoend}

As we did in the separated cases we also present here the temporal evolution of $n_{COM}$. We will begin with the case $\alpha_0\ll 1$ and then proceed with the opposite case.
\begin{figure}[ht]
	\centering
		\includegraphics[width=0.50\textwidth]{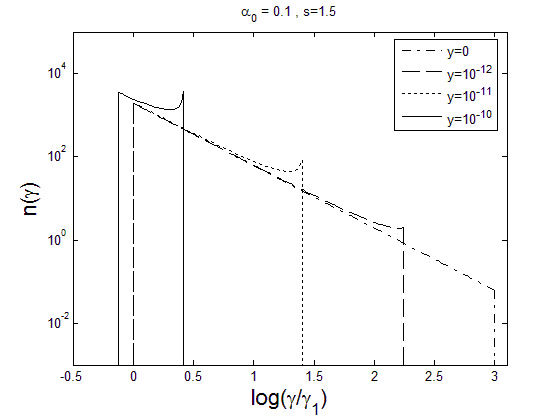}
	\caption{Temporal evolution of the combined electron number density for $s=1.5$ and $\alpha_0=0.1$ at four different times $y$ with the initial values $\gamma_1=10$ and $\gamma_2=10^4$.}
	\label{fig:COMs15a01}
\end{figure} \\
\begin{figure}[ht]
	\centering
		\includegraphics[width=0.50\textwidth]{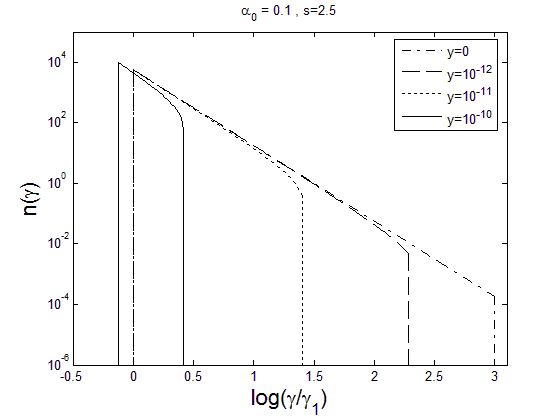}
	\caption{Temporal evolution of the combined electron number density for $s=2.5$ and $\alpha_0=0.1$ at four different times $y$ with the initial values $\gamma_1=10$ and $\gamma_2=10^4$.}
	\label{fig:COMs25a01}
\end{figure} \\
\begin{figure}[ht]
	\centering
		\includegraphics[width=0.50\textwidth]{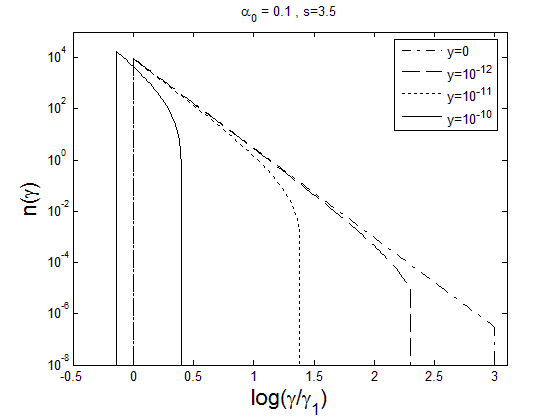}
	\caption{Temporal evolution of the combined electron number density for $s=3.5$ and $\alpha_0=0.1$ at four different times $y$ with the initial values $\gamma_1=10$ and $\gamma_2=10^4$.}
	\label{fig:COMs35a01}
\end{figure} \\
The results are indeed comparable to the results of section \ref{ssec:lc}. The width of the electron distribution shrinks linearly and similarly for all shown cases. The major difference is the pile-up for $s<2$ at the high-energy end of the distribution and that the spectra steepen for higher values of $s$. 
\begin{figure}[ht]
	\centering
		\includegraphics[width=0.50\textwidth]{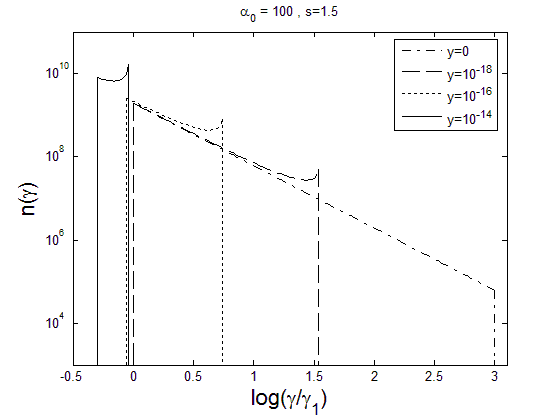}
	\caption{Temporal evolution of the combined electron number density for $s=1.5$ and $\alpha_0=100$ at four different times $y$ with the initial values $\gamma_1=10$ and $\gamma_2=10^4$.}
	\label{fig:COMs15a10}
\end{figure} \\
\begin{figure}[ht]
	\centering
		\includegraphics[width=0.50\textwidth]{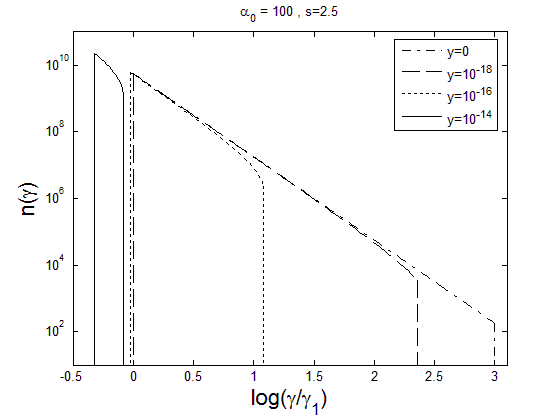}
	\caption{Temporal evolution of the combined electron number density for $s=2.5$ and $\alpha_0=100$ at four different times $y$ with the initial values $\gamma_1=10$ and $\gamma_2=10^4$.}
	\label{fig:COMs25a10}
\end{figure} \\
\begin{figure}[ht]
	\centering
		\includegraphics[width=0.50\textwidth]{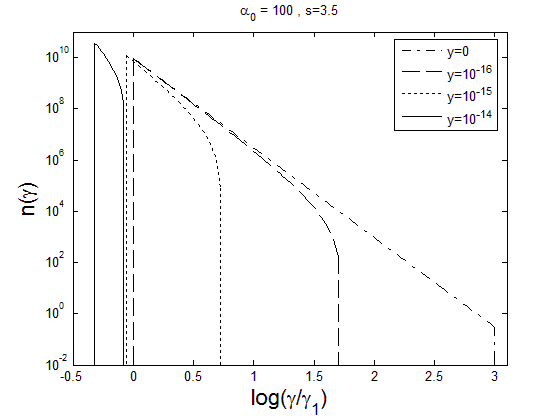}
	\caption{Temporal evolution of the combined electron number density for $s=3.5$ and $\alpha_0=100$ at four different times $y$ with the initial values $\gamma_1=10$ and $\gamma_2=10^4$.}
	\label{fig:COMs35a10}
\end{figure} \\
The important point of the plots \ref{fig:COMs15a10} - \ref{fig:COMs35a10} is that the cooling is much quicker than in the previous case for $\alpha_0\ll 1$. One should also notice the pile-up again at the high-energy end of the electron distribution for the hard spectrum, while for higher spectral indices the spectrum drops off.


\subsection{Temporal evolution of the cut-offs} \label{ssec:comteco}

\begin{figure}[ht]
	\centering
		\includegraphics[width=0.50\textwidth]{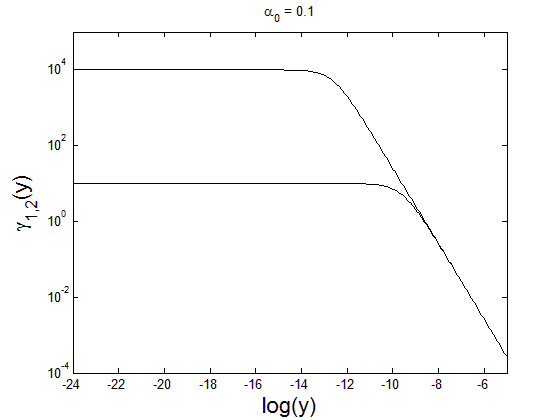}
	\caption{Temporal evolution of the cut-offs of the combined electron number density for $s=2.0$ and $\alpha_0=0.1$ for the initial values $\gamma_1=10$ and $\gamma_2=10^4$.}
	\label{fig:COMheavia01}
\end{figure}

\begin{figure}[ht]
	\centering
		\includegraphics[width=0.50\textwidth]{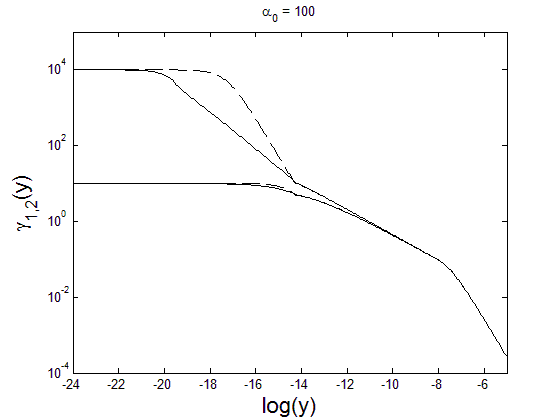}
	\caption{Temporal evolution of the cut-offs of the combined electron number density for $s=2.0$ (full curve), $s=3.5$ (dashed curve) and $\alpha_0=100$ for the initial values $\gamma_1=10$ and $\gamma_2=10^4$.}
	\label{fig:COMheavia10}
\end{figure}

In this section we want to verify the results of the last section, especially of the temporal behavior of the electron distribution. Therefore, we discuss the evolution of the cut-offs for three different cases of $s$ shown in the Figs. \ref{fig:COMheavia01} and \ref{fig:COMheavia10}.
 \\
 \\
Since for $\alpha_0\ll 1$ the behavior of the cut-offs is independent of $s$, we only plotted one case in Fig. \ref{fig:COMheavia01}. It is obvious that this is indeed very similar to the linear solution of section \ref{ssec:lc} (with the plot of the cut-offs in section \ref{ssec:teco}). \\
For $\alpha_0\gg 1$ we notice that the cooling begins much earlier than in the other case, with the hard case beginning even two orders of magnitude earlier in the example shown here than the soft case. We also notice that the soft case cools linearly in the beginning until it merges again with the already non-linearly cooled hard case (here, at roughly $y\approx 10^{-14}$). Interestingly, form that moment on both cases cool the same. When both begin to cool linearly again at late times (at about $y>10^{-8}$) they match the cooling of the case with $\alpha_0\ll 1$. \\
Obviously, the electron distributions of all cases undergo severe quenching resulting in the appearance of a $\delta$-function for later times. \\
Overall we can say that the behavior of the combined cooling is similar to the respective cases of the separated cooling scenarios. In order to stress the most important point once more, the non-linear case turns linear at late times as we expected (cf. section \ref{ssec:teco}).


\section{Linear synchrotron fluence} \label{sec:lsf}

In the next three sections we will derive the synchrotron fluence spectra, for which we need at first the respective intensity spectra. We use the monochromatic approximation of the synchrotron power introduced by Felten \& Morrison (1966). In general, the synchrotron intensity spectrum of a homogeneous, spherical, optically thin source is then given by
\be
I(\nu,t)=\frac{R}{4\pi}\frac{c\sigma_T B^2}{6\pi}\intl_{0}^{\infty}\td{\gamma}n(\gamma,t)\gamma^2\DELF{\nu-\nu_s\gamma^2} \label{intensityspec1} \eqc
\ee
where $\nu$ is the frequency of the emitted photon. The characteristic frequency is defined by $\nu_s=\frac{3eB}{4\pi mc^2}=4.2\cdot 10^{6} b\,\unit{Hz}$. \\
We can use the substitution rule for $\delta$-functions, providing $2\gamma\nu_s\DELF{\nu-\nu_s\gamma^2}=\DELF{\gamma-\sqrt{\nu/\nu_s}}$. Inserting this into Eq. (\ref{intensityspec1}) results in:
\be
I(\nu,t)=\frac{R}{4\pi}\frac{c\sigma_T B^2}{12\pi\nu_s}\intl_{0}^{\infty}\td{\gamma}n(\gamma,t)\gamma\DELF{\gamma-\sqrt{\frac{\nu}{\nu_s}}} \label{intensityspec2} \eqd
\ee
If we now apply this to the linear electron distribution Eq. (\ref{lineldis}), define the normalized frequency $f=\frac{\nu}{\nu_s\gamma_1^2}$, and a new time variable $\chi=\gamma_1 t=\frac{\gamma_1}{A_0}y$, we obtain
\be
I_{SYN}(f,\chi) = \frac{Rq_0 c\sigma_T B^2}{48\pi^2\nu_s} \gamma_1^{1-s} f^{\frac{1-s}{2}} \left( 1-D_0 f^{1/2} \chi \right)^{s-2} \nonumber \\
\times \HSF{f- \frac{1}{(1+D_0 \chi)^2}} \HSF{\frac{1}{(\ggg+D_0 \chi)^2} -f} \label{isynsyn2} \eqd
\ee
The fluence is the time-integrated intensity spectrum. Setting an arbitrary upper limit would result in the fractional fluence, which shows the fluence after some time. Here we treat, however, only the total fluence with an upper limit that is infinite. Thus,
\be
F_{SYN}(f) = \frac{Rq_0c\sigma_TB^2}{48\pi^2\nu_s\gamma_1^s}f^{\frac{1-s}{2}}\intl_{0}^{\infty}\td{\chi}\left( 1-D_0f^{1/2}\chi \right)^{s-2} \nonumber \\
\times \HSF{f- \frac{1}{(1+D_0 \chi)^2}} \HSF{\frac{1}{(\ggg+D_0 \chi)^2} -f} \label{fsynsyn1} \eqd
\ee
Introducing the variable $\Omega=D_0f^{1/2}\chi$ we yield
\be
F_{SYN}(f)=\frac{Rq_0c\sigma_TB^2}{48\pi^2\nu_s\gamma_1^sf^{s/2}D_0}\intl_{0}^{\infty}\td{\Omega}(1-\Omega)^{s-2} \nonumber \\
\times \HSF{\Omega- (1-f^{1/2})} \HSF{(1-\ggg f^{1/2}) -\Omega} \label{fsynsyn2} \eqd
\ee
The result of this integral gives two different frequency regimes, and can be written as
\be
F_{SYN}(f)=N_l \nonumber \\
\times \left\{ \begin{array}{ll}
f^{-1/2}\left( 1-g_2^{1-s} \right) & \,\, ,\, f\leq 1 \\
f^{-s/2}\left( 1-\left( \frac{f}{g_2^2} \right)^{\frac{s-1}{2}} \right) & \,\, ,\, 1\leq f\leq g_2^2
  \end{array} \right. \label{fsynsyn} \eqc
\ee
with $N_l=\frac{Rq_0c\sigma_TB^2}{48\pi^2\nu_s\gamma_1^sD_0(s-1)}$. \\
This is the well known result that has been obtained by Schlickeiser \& Lerche (2008), too. The important points of this result are that the fluence is independent of the spectral index $s$ of the electron injection rate for frequencies smaller than unity following a power-law with a spectral index $\alpha_{SYN}=1/2$. At $f=1$ the spectrum exhibits a break to a steeper spectrum with $\Delta\alpha_{SYN}=\frac{s-1}{2}$ depending on $s$. The spectrum is cut off at $f=g_2^2$.
\begin{figure}[ht]
	\centering
		\includegraphics[width=0.50\textwidth]{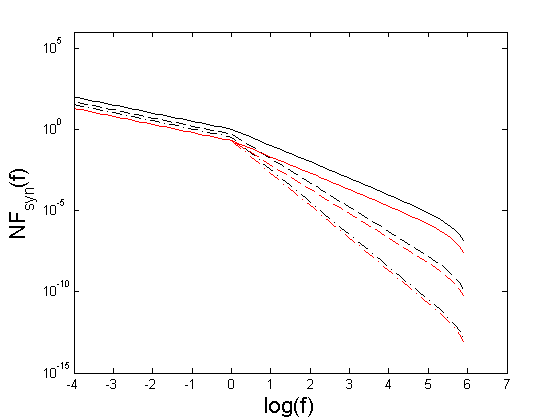}
	\caption{$NF_{SYN}$ as a function of $f$ for three cases of $s$ (\textit{Full:} $s=2$, \textit{Dashed:} $s=3$, \textit{Dot-dashed:} $s=4$) and for $g_2=10^3$. The red lines indicate the analytical solution Eq. (\ref{fsynsyn}) with an offset of $10^{-0.7}$.}
	\label{fig:pNFg23}
\end{figure}\\
In Fig. \ref{fig:pNFg23} we show the numerical solution of the fluence spectrum for which we normalized the fluence by $NF_{SYN}=F_{SYN} / N_l$. We plotted the result for $g_2=10^3$ and three values of $s$. We also plotted the analytical results, but with a small offset in order to highlight them. \\
As one can see, the numerical and the analytical result match each other rather well, which is not so surprising, since we introduced no strong approximations during the analytical work.


\section{Nonlinear synchrotron fluence} \label{sec:sfs}

In order to obtain the intensity spectrum and the fluence of the nonlinear case we have to insert the nonlinear electron distribution Eq. (\ref{nleldis}) into Eq. (\ref{intensityspec2}), yielding
\be
I_{SSC}(f,x) = \frac{Rq_0 c\sigma_T B^2}{48\pi^2\nu_s} \gamma_1^{1-s} f^{\frac{1-s}{2}} \left( 1-f^{1/2} x \right)^{s-2} \nonumber \\
\times \HSF{f- \frac{1}{(1+x)^2}} \HSF{\frac{1}{(\ggg+x)^2} -f} \label{isynssc1} \eqd
\ee
Integrating this with respect to time yields the total fluence:
\be
F_{SSC}(f) = \frac{Rq_0c\sigma_TB^2}{48\pi^2\nu_sA_0\gamma_1^s} f^{-s/2}\intl_{0}^{\infty}\td{\omega}U^{-1}\left(\frac{\omega}{f^{1/2}}\right)  \nonumber \\
\times (1-\omega)^{s-2}\HSF{\omega- (1-f^{1/2})} \nonumber \\
\times \HSF{(1-\ggg f^{1/2}) -\omega} \label{fsynssc2} \eqc
\ee
for which we used the substitution $\omega=f^{1/2}x$. \\ 
Since we must take care of the spectral index $s$, we split the ongoing discussion into two parts for each case of the spectral index.


\subsection{Large spectral index} \label{ssec:lsi}

We begin with the spectral index being larger than $3$. Then, the first integral we have to calculate is for the case $0\leq\omega\leq\ggg f^{1/2}$, where $U(\omega/f^{1/2})=U_0(1-g_2^{3-s})$, yielding
\be
F_{SSC}(f) = N_0\frac{f^{-s/2}}{1-g_2^{3-s}} \intl_{0}^{\ggg f^{1/2}}\td{\omega}(1-\omega)^{s-2} \nonumber \\
\times\HSF{\omega-(1-f^{1/2})}\HSF{(1-\ggg f^{1/2})-\omega} \label{fsynsscs3a1} \eqc
\ee
where we set $N_0=\frac{Rq_0c\sigma_TB^2}{48\pi^2\nu_sA_0\gamma_1^sU_0}$. \\
The integration can be easily performed giving, according to the Heaviside functions, solutions for two regimes of the normalized frequency:
\be
F_{SSC}(f) = N_0 \frac{f^{-s/2}}{1-g_2^{3-s}}\frac{1}{s-1} \nonumber \\
\times \left\{ \begin{array}{ll} 
1-\left( 1-\left( \frac{f}{g_2^2} \right)^{1/2} \right)^{s-1} & \,\, ,\,1\leq f\leq \frac{1}{4}g_2^2 \\
1-\left( \frac{f}{g_2^2} \right)^{\frac{s-1}{2}} & \,\, ,\, \frac{1}{4}g_2^2\leq f\leq g_2^2
\end{array} \right. \label{fsynsscs3a2} \eqd
\ee 
The intermediate $\omega$-range ($\ggg f^{1/2}\leq\omega\leq f^{1/2}$) is the next case we have to consider. Here 
\be
U(\omega)=U_0\left[ 1-\frac{2(\omega/f^{1/2})^{s-3}}{s-1}-\frac{s-3}{s-1}\frac{g_2^{1-s}}{(\omega/f^{1/2})^2} \right] \eqd
\ee
Approximating this with the leading term, which is valid for most parts of the time period and for spectral indices not too close to $3$ (cf. Fig. \ref{fig:Uintes3a} in appendix \ref{sec:appx}), the integral becomes
\be
F_{SSC}(f) = N_0 f^{-s/2} \intl_{\ggg f^{1/2}}^{f^{1/2}}\td{\omega}(1-\omega)^{s-2} \nonumber \\
\times \HSF{\omega-(1-f^{1/2})}\HSF{(1-\ggg f^{1/2})-\omega} \label{fsynsscs3b1} \eqd
\ee
This is again easily solved for two regimes of $f$ obtained from the Heaviside functions and indicated below. We find
\be
F_{SSC}(f) = N_0 f^{-s/2}\frac{1}{s-1} \nonumber \\
\times \left\{ \begin{array}{ll} 
f^{\frac{s-1}{2}}-\left(1-f^{1/2}\right)^{s-1} & \,\, ,\,1/4\leq f\leq 1 \\
\left( 1-\ggg f^{1/2} \right)^{s-1} - \left( \frac{f}{g_2^2} \right)^{\frac{s-1}{2}} & \,\, ,\, 1\leq f\leq \frac{1}{4}g_2^2
\end{array} \right. \label{fsynsscs3b2} \eqd
\ee
The third time regime $\omega\geq f^{1/2}$ requires $U=U_0\frac{s-3}{s-1}(1-g_2^{1-s})f\omega^{-2}$, for which the fluence becomes
\be
F_{SSC}(f) = N_0 \frac{s-1}{s-3}(1-g_2^{1-s})^{-1}f^{-\frac{s+2}{2}} \intl_{f^{1/2}}^{\infty}\td{\omega}\omega^2 \nonumber \\
\times (1-\omega)^{s-2}\HSF{\omega-(1-f^{1/2})} \nonumber \\
\times \HSF{(1-\ggg f^{1/2})-\omega} \label{fsynsscs3c1} \eqc
\ee
The primitive of the integral can be obtained by two integration by parts, yielding for arbitrary limits $a$ and $b$, which need to be specified by the Heaviside functions later,
\be
\mathrm{INT} = \intl_{a}^{b}\td{\omega}\omega^2(1-\omega)^{s-2} \nonumber \\
=\left[ -\frac{\omega^2}{s-1}(1-\omega)^{s-1} \right]_a^b + \left[ -\frac{2\omega}{s^2-s}(1-\omega)^s \right]_a^b \nonumber \\
+ \left[ -\frac{2}{s^3-s}(1-\omega)^{s+1} \right]_a^b \label{int3prim} \eqd
\ee
For $f\leq 1/4$ the limits become $a=1-f^{1/2}$ and $b=(1-\ggg f^{1/2})$. Assuming $g_2\gg 1$ we can neglect the negative terms, which are the terms where we inserted the upper limit. Since $f\leq 1/4$ we see that the second and third of the remaining terms are also much reduced compared to the first one. Thus, we approximate the integral with
\be
\mathrm{INT} \approx \frac{(1-f^{1/2})^2}{s-1}f^{\frac{s-1}{2}} \label{int3alpha2} \eqd
\ee
From the Heaviside functions of Eq. (\ref{fsynsscs3c1}) we obtain another frequency regime located between $1/4\leq f\leq 1$. The limits in this case become $a=f^{1/2}$ and $b=1-\ggg f^{1/2}$. Using the same approximation as above, for which we note however that these approximations are not as well fitting as in the previous case, we find
\be 
\mathrm{INT} \approx \frac{f}{s-1}\left( 1-f^{1/2} \right)^{s-1} \label{int3beta2} \eqd
\ee
Thus, we yield in the late time case for the fluence
\be
F_{SSC}(f) = N_0 \frac{s-1}{s-3}(1-g_2^{1-s})^{-1}f^{-\frac{s+2}{2}} \nonumber \\
\times \left\{ \begin{array}{ll} 
\frac{(1-f^{1/2})^2}{s-1}f^{\frac{s-1}{2}} & \,\, ,\, f\leq 1/4 \\
\frac{f}{s-1}\left( 1-f^{1/2} \right)^{s-1} & \,\, ,\, 1/4\leq f\leq 1
\end{array} \right. \label{fsynsscs3c2} \eqd
\ee
Collecting terms we find for the total fluence with a large spectral index $s>3$
\be
F_{SSC}(f)=N_0\left[ F_1(f< 1/4)+F_2(1/4\leq f< 1) \right. \nonumber \\
					\left. +F_3(1\leq f<\frac{1}{4}g_2^2)+F_4(\frac{1}{4}g_2^2\leq f\leq g_2^2) \right] \eqc
\ee
with the terms
\be
F_1 &=&  \frac{(1-f^{1/2})^2}{(s-3)(1-g_2^{1-s})}f^{-3/2} \label{fsynsscs3p1} \\
F_2 &=&  \frac{f^{-s/2}}{s-1}\left[ f^{\frac{s-1}{2}}-(1-f^{1/2})^{s-1} \right] \nonumber \\
    & & +\frac{f^{-s/2}}{(s-3)(1-g_2^{1-s})}(1-f^{1/2})^{s-1} \label{fsynsscs3p2} \\
F_3 &=&  \frac{f^{-s/2}}{(s-1)(1-g_2^{3-s})}  \left[ 1-\left( 1-\left( \frac{f}{g_2^2} \right)^{1/2} \right)^{s-1} \right] \nonumber \\
    & & +\frac{f^{-s/2}}{s-1} \left[ \left( 1-\left( \frac{f}{g_2^2} \right)^{1/2} \right)^{s-1} - \left( \frac{f}{g_2^2} 			   			\right)^{\frac{s-1}{2}} \right] \nonumber \\
    &\approx& \frac{f^{-s/2}}{s-1} \left[ 1-\left( \frac{f}{g_2^2} \right)^{\frac{s-1}{2}} \right] \label{fsynsscs3p3} \\
F_4 &=&  \frac{f^{-s/2}}{(s-1)(1-g_2^{3-s})} \left[ 1-\left( \frac{f}{g_2^2} \right)^{\frac{s-1}{2}} \right] \label{fsynsscs3p4} \eqd
\ee
In the third part we approximated again for $g_2\gg 1$. If we do the same in the fourth part, the third and fourth part would be the same and could be merged. This does not change much to the overall result, which should be discussed briefly. \\
The fluence behaves like a power-law with a constant spectral index $\alpha_{SSC}=3/2$ for $f\leq1$, independently of $s$, which is steeper than in the case of linear cooling. At $f\approx 1$ the spectrum exhibits a spectral break to a steeper power-law with a change in the spectral index of $\Delta\alpha_{SSC}=(s-3)/2$. At a frequency $f=g_2^2$ the spectrum is cut off at the high energy end.
\begin{figure}[ht]
	\centering
		\includegraphics[width=0.50\textwidth]{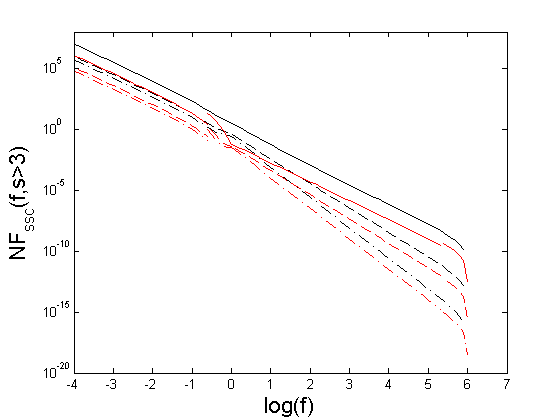}
	\caption{$NF_{SSC}$ as a function of $f$ for three cases of $s>3$ (\textit{Full:} $s=3.1$, \textit{Dashed:} $s=4$, \textit{Dot-dashed:} $s=5$) and for $g_2=10^3$. The red lines indicate the analytical solution Eqs. (\ref{fsynsscs3p1}) - (\ref{fsynsscs3p4}) with an offset of $10^{-0.9}$.}
	\label{fig:pNFs3g23}
\end{figure}\\
In Fig. \ref{fig:pNFs3g23} we show the numerical solution of the fluence spectrum for which we normalized the fluence by $NF_{SSC}=F_{SSC} / N_0$. We plotted the result for $g_2=10^3$ and three values of $s$. We also plotted the analytical results, but with a small offset in order to highlight them. \\
Obviously for most parts of $f$ the analytical result and the numerical result match each other. However, for $1/4\leq f<1$ there is a discrepancy between the results, which is also quite obvious, since the analytical result is not continuous at $f=1/4$. We already indicated during the analytical calculations that the approximation for the second part (Eq. (\ref{fsynsscs3p2})) could be invalid. One can also see that the approximation in the third part (Eq. (\ref{fsynsscs3p3})) becomes less valid for $s\rightarrow 3$. Thus, the numerical result confirms the expectations we mentioned during the derivation of the analytical result.


\subsection{Small spectral index} \label{ssec:ssi}

Now, we want to deal with the fluence for electron spectral indices $1<s<3$. As a matter of fact, the steps are quite similar to the case before, but nonetheless we will repeat them here. \\
We begin again with the case $0\leq\omega\leq\ggg f^{1/2}$, where $U(\omega/f^{1/2})=U_1(g_2^{3-s}-1)$:
\be
F_{SSC}(f) = N_1\frac{f^{-s/2}}{g_2^{3-s}-1} \intl_{0}^{\ggg f^{1/2}}\td{\omega}(1-\omega)^{s-2} \nonumber \\
\times \HSF{\omega-(1-f^{1/2})}\HSF{(1-\ggg f^{1/2})-\omega} \label{fsynssc1s3a1} \eqc
\ee
with the definition $N_1=\frac{Rq_0c\sigma_TB^2}{48\pi^2\nu_sA_0\gamma_1^sU_1}$.
Since the integral is easily obtained once more, for small $\omega$ the fluence becomes
\be
F_{SSC}(f) = N_1 \frac{f^{-s/2}}{g_2^{3-s}-1}\frac{1}{s-1} \nonumber \\
\times \left\{ \begin{array}{ll} 
1-\left( 1-\left( \frac{f}{g_2^2} \right)^{1/2} \right)^{s-1} & \,\, ,\,1\leq f\leq \frac{1}{4}g_2^2 \\
1-\left( \frac{f}{g_2^2} \right)^{\frac{s-1}{2}} & \,\, ,\, \frac{1}{4}g_2^2\leq f\leq g_2^2
\end{array} \right. \label{fsynssc1s3a2} \eqc
\ee 
where the Heaviside functions of Eq. (\ref{fsynssc1s3a1}) defined the frequency regimes. \\
The intermediate $\omega$-range ($\ggg f^{1/2}\leq\omega\leq f^{1/2}$) is the next case we turn our attention to. Here 
\be
U(\omega)=U_1\left[ \frac{2(\omega/f^{1/2})^{s-3}}{s-1}-1+\frac{3-s}{s-1}\frac{g_2^{1-s}}{(\omega/f^{1/2})^2} \right] \eqd
\ee
Taking into account only the leading term of $U(\omega)$ (the validity of this approximation is shown in Fig. \ref{fig:Uinte1s3a} in appendix \ref{sec:appx}), as we did in the same case for the steep electron spectrum, we obtain 
\be
F_{SSC}(f) = N_1 f^{-s/2} \frac{s-1}{2f^{\frac{3-s}{2}}}\intl_{\ggg f^{1/2}}^{f^{1/2}}\td{\omega}\omega^{3-s}(1-\omega)^{s-2} \nonumber \\
\times \HSF{\omega-(1-f^{1/2})}\HSF{(1-\ggg f^{1/2})-\omega} \label{fsynssc1s3b2} \eqc
\ee
This integral can be expressed in terms of the hypergeometric function, yielding no simple solution. Thus, for intermediate frequencies we cannot present an analytical solution. \\
Finally, we have to turn our attention to the late time regime, which means $\omega\geq f^{1/2}$. Here $U=U_1\frac{3-s}{s-1}\frac{f}{\omega^2}(1-g_2^{1-s})$, for which the fluence becomes
\be
F_{SSC}(f) = N_1 \frac{f^{-\frac{s+2}{2}}}{1-g_2^{1-s}}\frac{s-1}{3-s} \intl_{f^{1/2}}^{\infty}\td{\omega}\omega^2(1-\omega)^{s-2} \nonumber \\
\times \HSF{\omega-(1-f^{1/2})}\HSF{(1-\ggg f^{1/2})-\omega} \label{fsynssc1s3c1} \eqd
\ee
This is almost the same integral that we solved in the case of $s>3$, thus we can use the same primitive (Eq. (\ref{int3prim})). In fact, the limits and the frequency regimes are the same as in the case of steeply injected electrons. Therefore, using the same approximations as we did in that case, we obtain the solution for $f\leq 1/4$
\be
\mathrm{INT} \approx \frac{(1-f^{1/2})^2}{s-1}f^{\frac{s-1}{2}} \label{int6alpha2} \eqc
\ee
and for $1/4\leq f\leq 1$
\be 
\mathrm{INT} \approx \frac{f}{s-1}\left( 1-f^{1/2} \right)^{s-1} \label{int6beta2} \eqd
\ee
Collecting terms, we find the fluence for large $\omega$:
\be
F_{SSC}(f) = N_1 \frac{s-1}{3-s}(1-g_2^{1-s})^{-1}f^{-\frac{s+2}{2}} \nonumber \\
\times \left\{ \begin{array}{ll} 
\frac{(1-f^{1/2})^2}{s-1}f^{\frac{s-1}{2}} & \,\, ,\, f\leq 1/4 \\
\frac{f}{s-1}\left( 1-f^{1/2} \right)^{s-1} & \,\, ,\, 1/4\leq f\leq 1
\end{array} \right. \label{fsynssc1s3c2} \eqd
\ee
Now, we have performed all necessary steps to calculate the synchrotron fluence of SSC cooled electrons that are injected into the jet with a spectral index $1<s<3$. Therefore,
\be
F_{SSC}(f) = N_1 \left[ F_5(f< 1/4)+F_6(\frac{1}{4}g_2^2\leq f\leq g_2^2) \right] \eqc
\ee
with
\be
F_5 &=&  \frac{(1-f^{1/2})^2}{(3-s)(1-g_2^{1-s})}f^{-3/2} \label{fsynssc1s3p1} \\
F_6 &=&  \frac{f^{-s/2}}{(s-1)(g_2^{3-s}-1)} \left[ 1-\left( \frac{f}{g_2^2} \right)^{\frac{s-1}{2}} \right] \label{fsynssc1s3p4} \eqd
\ee
We do not present the results for the intermediate frequency regime $1/4<f\leq1/4g_2^2$, because this is the part where the intermediate $\omega$-regime is also important, if not more important than the contributions of the early and late $\omega$-regimes. And for that part we did not obtain an analytical solution. \\
However, the results in the other frequency parts are noteworthy. The fluence for flat injected electron spectra shows similar characteristics as the fluence for steep injected electrons. We find, again, that the spectral index does not depend on $s$ for frequencies $f<1/4$ and has the same value $\alpha_{SSC}=3/2$. There is also a cut-off for high frequencies at $f=g_2^2$. 
\begin{figure}[ht]
	\centering
		\includegraphics[width=0.50\textwidth]{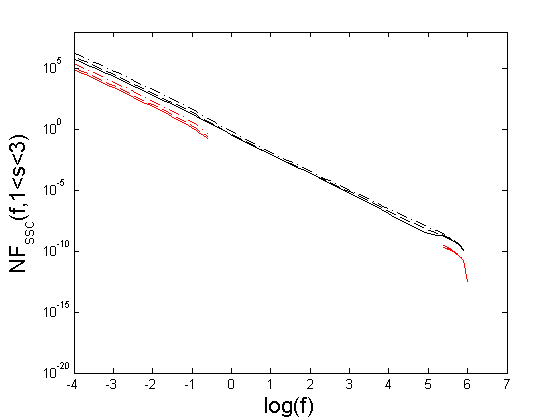}
	\caption{$NF_{SSC}$ as a function of $f$ for three cases of $s>3$ (\textit{Full:} $s=3.1$, \textit{Dashed:} $s=4$, \textit{Dot-dashed:} $s=5$) and for $g_2=10^3$. The red lines indicate the analytical solution Eqs. (\ref{fsynssc1s3p1}) and (\ref{fsynssc1s3p4}) with an offset of $10^{-0.9}$.}
	\label{fig:pNF1s3g23}
\end{figure}\\
In Fig. \ref{fig:pNF1s3g23} we show the numerical solution of the fluence spectrum for which we normalized the fluence by $NF_{SSC}=F_{SSC} / N_1$. We plotted the result for $g_2=10^3$ and three values of $s$. We also plotted the analytical results, but with a small offset in order to highlight them. \\
As one can see, the numerical and the analytical solutions agree very well for the presented frequency intervals. Interestingly, there is no spectral break in the numerical result at around $f=1$, different from the behavior of steeply injected electrons. The spectrum is cut off at $f=g_2^2$ as expected. \\
It is interesting that the spectrum does not exhibit a break at $f=1$, implying that the spectrum for $f>1$ does not depend strongly on $s$, either. It is quite obvious that the offset between the numerical and the analytical solution does not change from the low frequency regime to the high frequency regime near the cut-off. The curves of the analytical plot at the cut-off are also equal for all cases, similar to the numerical result. Especially this last point is different from the steep injection case, where the magnitudes of the curves differ significantly from case to case. These are hints for the validity of the numerical solution exhibiting no break, which is an important difference from the steep injection case.   


\section{Synchrotron fluence of combined cooled electrons} \label{sec:sfcom}

We will, now, calculate the synchrotron fluence spectrum of electrons undergoing the combined cooling. \\
As before, we begin with the intensity by inserting Eq. (\ref{comeldis}) into Eq. (\ref{intensityspec2}). With the same definition for the normalized frequency $f$ we obtain
\be
I_{COM}(f,\xc)=\frac{q_0Rc\sigma_TB^2}{48\pi^2\nu_s}\gamma_1^{1-s} f^{\frac{1-s}{2}} \left( 1-f^{1/2}\xc \right)^{s-2} \nonumber \\
\times \HSF{\frac{1}{(\ggg+\xc)^2}-f}\HSF{f-\frac{1}{(1+\xc)^2}} \label{intspeccom} \eqd
\ee
The integration with respect to time yields the synchrotron fluence:
\be
\Fc(f)=\intl_0^{\infty}\td{t}I_{COM}(f,\xc)=\frac{1}{A_0}\intl_0^{\infty}\td{y}I_{COM}(f,\xc(y)) \nonumber \\
=\Fa f^{\frac{1-s}{2}}\intl_0^{\infty}\td{y} \left( 1-f^{1/2}\xc(y) \right)^{s-2} \nonumber \\
\times \HSF{\frac{1}{(\ggg+\xc(y))^2}-f}\HSF{f-\frac{1}{(1+\xc(y))^2}} \label{fcom} \eqc
\ee
with the definition $\Fa=\frac{q_0Rc\sigma_TB^2}{48\pi^2\nu_sA_0\gamma_1^{s-1}}$. \\
It is advantageous in this case to use $y$ as the integration variable instead of $\xc$, as we did in the other cases, because we already made a lot of approximations while calculating $\xc(y)$ for the discussions in section \ref{sec:ccend}, and it is unnecessary to repeat these steps again. \\
Using $\xc(y)$ form Eqs. (\ref{comtimevabsol1}) - (\ref{comtimevabsol4}) we will in each of the following integrations define the new variable $\Psi:=f^{1/2}\xc(y)$.


\subsection{Large spectral index, small injection parameter} \label{ssec:lssi}

Here we discuss the case of $s>3$ and $\alpha_0\ll 1$. \\
The first regime we have to consider is for $0\leq\Psi\leq\Psi_2$, with $\Psi_2:=\Psi(y_2)=f^{1/2}$. With $\td{\Psi}=f^{1/2}\gamma_1(K_0+U_0)\td{y}$ we find
\be
\Fc(f)=\Fa\frac{f^{-s/2}}{\gamma_1(K_0+U_0)}\intl_0^{\Psi_2}\td{\Psi}(1-\Psi)^{s-2} \nonumber \\
\times \HSF{(1-f^{1/2}\ggg)-\Psi}\HSF{\Psi-(1-f^{1/2})} \nonumber \\
=\Fa \frac{f^{-s/2}}{\gamma_1(K_0+U_0)(s-1)} \left[ -(1-\Psi)^{s-1} \right]_{a}^{b} \label{fs3a01a} \eqc
\ee
where $a$ and $b$ are defined by the limits of the integral and the Heaviside functions. For $f<1$ we find $a=1-f^{1/2}$ and $b=f^{1/2}$, resulting in
\be
\Fc(f<1)=\Fa \frac{f^{-1/2}}{\gamma_1(K_0+U_0)(s-1)} \nonumber \\
\times \left[ 1-(f^{-1/2}-1)^{s-1} \right] \label{fs3a01a1} \eqc
\ee
This becomes negative for $f<1/4$ which is not physically possible, and, therefore, this solution does not contribute for lower frequencies. \\
For $f>1$ we obtain the limits $a=0$ and $b=1-f^{1/2}\ggg$, resulting in
\be
\Fc(f>1)=\Fa \frac{f^{-s/2}}{\gamma_1(K_0+U_0)(s-1)} \nonumber \\ 
\times \left[ 1-\left( \frac{f}{g_2^2} \right)^{\frac{s-1}{2}} \right] \label{fs3a01a2} \eqd
\ee
Obviously, this is cut off at $f=g_2^2$. \\
For $\Psi>\Psi_2$ we find $\td{\Psi}=f^{1/2}\gamma_1K_0\td{y}$, leading to
\be
\Fc(f)=\Fa \frac{f^{-s/2}}{\gamma_1K_0}\intl_{\Psi_2}^{\infty}\td{\Psi}(1-\Psi)^{s-2} \nonumber \\
\times \HSF{(1-f^{1/2}\ggg)-\Psi}\HSF{\Psi-(1-f^{1/2})} \nonumber \\
=\Fa \frac{f^{-s/2}}{\gamma_1K_0(s-1)} \left[ -(1-\Psi)^{s-1} \right]_a^b \label{fs3a01b} \eqd
\ee
We find for $f<1/4$ that $a=1-f^{1/2}$ and $b=1-f^{1/2}\ggg$, yielding
\be
\Fc(f<1/4)=\Fa\frac{f^{-1/2}}{\gamma_1K_0(s-1)} \left[ 1-g_2^{1-s} \right] \label{fs3a01b1} \eqd
\ee
For $f>1/4$ we obtain for $a=f^{1/2}$ and $b=1-f^{1/2}\ggg$
\be
\Fc(f>1/4)=\Fa\frac{f^{-1/2}}{\gamma_1K_0(s-1)} \nonumber \\
\times \left[ (f^{-1/2}-1)^{s-1}-g_2^{1-s} \right] \label{fs3a01b2} \eqc
\ee
which becomes negative for $f>1$. \\
Collecting terms, we obtain the total fluence 
\be
\Fc(s>3,\alpha_0\ll 1)=\Fa \left[ F_c^1(f<1/4)\right. \nonumber \\
\left.+F_c^2(1/4<f<1)+F_c^3(1<f<g_2^2) \right] \label{fs3a01sol} \eqd
\ee
with
\be
F_c^1 &=& \frac{f^{-1/2}}{\gamma_1K_0(s-1)} \left[ 1-g_2^{1-s} \right] \\
F_c^2 &=& \frac{f^{-1/2}}{\gamma_1(s-1)} \left\{ \frac{1}{K_0}\left[ (f^{-1/2}-1)^{s-1}-g_2^{1-s} \right] \right. \nonumber \\
& & \left.+ \frac{1}{K_0+U_0}\left[ 1-(f^{-1/2}-1)^{s-1} \right] \right\} \\
F_c^3 &=& \frac{f^{-s/2}}{\gamma_1(K_0+U_0)(s-1)} \left[ 1-\left( \frac{f}{g_2^2} \right)^{\frac{s-1}{2}} \right] \eqd
\ee
This is the expected result. Since the cooling is dominated by the linear part, the solution is comparable to the linear fluence of section \ref{sec:lsf} representing a broken power-law with a constant spectral index ($\beta_1=1/2$) for frequencies below unity, and a steepening of $\Delta\beta_1=(s-1)/2$ for $f\geq 1$. The spectrum is cut off at $f=g_2^2$.


\subsection{Small spectral index, small injection parameter} \label{ssec:sssi}

We continue with the case of $1<s<3$ and $\alpha_0\ll 1$. \\
The first part is $0\leq\Psi\leq\Psis_1$, with $\Psis_1:=\Psi(\ys_1)=f^{1/2}\ggg$. In this case, we find $\td{\Psi}=f^{1/2}\gamma_1(K_0+U_1g_2^{3-s})\td{y}$ leading to
\be
\Fc(f)=\Fa\frac{f^{-s/2}}{\gamma_1(K_0+U_1g_2^{3-s})}\intl_0^{\Psis_1}\td{\Psi}(1-\Psi)^{s-2} \nonumber \\
\times \HSF{(1-f^{1/2}\ggg)-\Psi}\HSF{\Psi-(1-f^{1/2})} \nonumber \\
=\Fa \frac{f^{-s/2}}{\gamma_1(K_0+U_1g_2^{3-s})(s-1)} \left[ -(1-\Psi)^{s-1} \right]_{a}^{b} \label{f1s3a01a} \eqd
\ee
With the help of the integration limits and the Heaviside functions we determine $a$ and $b$ for three different regimes of $f$. For $f<1$ $a=1-f^{1/2}$ and $b=f^{1/2}\ggg$ for which the fluence becomes
\be
\Fc(f<1)\approx \Fa \frac{f^{-1/2}}{\gamma_1(K_0+U_1g_2^{3-s})(s-1)} \nonumber \\
\times \left[ 1-f^{\frac{1-s}{2}} \right] \label{f1s3a01a1} \eqc
\ee
Since $f<1$ and $s>1$, this is always negative, which means, in turn, that this part does not contribute to the overall result. \\
For $1<f<1/4g_2^2$ we obtain the limits $a=0$ and $b=f^{1/2}\ggg$, resulting in
\be
\Fc(1<f<1/4g_2^2)=\Fa \frac{f^{-s/2}}{\gamma_1(K_0+U_1g_2^{3-s})(s-1)} \nonumber \\
\times \left[ 1-\left( 1-\left( \frac{f}{g_2^2} \right)^{1/2} \right)^{s-1} \right] \label{f1s3a01a2} \eqd
\ee
The third interval is for $f>1/4g_2^2$ with $a=0$ and $b=1-f^{1/2}\ggg$. Thus,
\be
\Fc(f>1/4g_2^2)=\Fa \frac{f^{-s/2}}{\gamma_1(K_0+U_1g_2^{3-s})(s-1)} \nonumber \\
\times \left[ 1-\left( \frac{f}{g_2^2} \right)^{\frac{s-1}{2}} \right] \label{f1s3a01a3} \eqd
\ee
This is, again, cut off at $f=g_2^2$. \\
For $\Psis_1<\Psi<\Psis_2$, where $\Psis_2=\Psi(\ys_2)=f^{1/2}$, the differential becomes $\td{\Psi}=f^{1/2}\frac{2\gamma_1U_1\beta}{(s-1)\alpha_0^2}\td{y}$, leading to
\be
\Fc(f)=\Fa \frac{f^{-s/2}(s-1)\alpha_0^2}{2\gamma_1U_1\beta}\intl_{\Psis_1}^{\Psis_2}\td{\Psi}(1-\Psi)^{s-2} \nonumber \\
\times \HSF{(1-f^{1/2}\ggg)-\Psi}\HSF{\Psi-(1-f^{1/2})} \nonumber \\
=\Fa \frac{f^{-s/2}\alpha_0^2}{2\gamma_1U_1\beta} \left[ -(1-\Psi)^{s-1} \right]_a^b \label{f1s3a01b} \eqd
\ee
For small frequencies $f<1$ the limits become $a=1-f^{1/2}$ and $b=f^{1/2}$, resulting in
\be
\Fc(f<1)=\Fa\frac{f^{-1/2}\alpha_0^2}{2\gamma_1U_1\beta} \left[ 1-\left( f^{-1/2}-1 \right)^{s-1} \right] \label{f1s3a01b1} \eqd
\ee
This turns negative for $f<1/4$ \\
In the case $f>1$ we obtain $a=f^{1/2}\ggg$ and $b=1-f^{1/2}\ggg$. Therefore,
\be
\Fc(f>1)=\Fa\frac{f^{-1/2}\alpha_0^2}{2\gamma_1U_1\beta} \nonumber \\
\times \left[ \left( 1-\left( \frac{f}{g_2^2} \right)^{1/2} \right)^{s-1}-\left( \frac{f}{g_2^2} \right)^{\frac{s-1}{2}} \right] \label{f1s3a01b2} \eqc
\ee
which does also not contribute for $f>1/4g_2^2$, since $\Fc<0$ in that part. \\
For $\Psi>\Psis_2$ we find $\td{\Psi}=f^{1/2}\gamma_1K_0\td{y}$, leading to
\be
\Fc(f)=\Fa \frac{f^{-s/2}}{\gamma_1K_0}\intl_{\Psis_2}^{\infty}\td{\Psi}(1-\Psi)^{s-2} \nonumber \\
\times \HSF{(1-f^{1/2}\ggg)-\Psi}\HSF{\Psi-(1-f^{1/2})} \nonumber \\
=\Fa \frac{f^{-s/2}}{\gamma_1K_0(s-1)} \left[ -(1-\Psi)^{s-1} \right]_a^b \label{f1s3a01c} \eqd
\ee
The results are obviously similar to the steep injection case, so that for $f<1/4$ $a=1-f^{1/2}$ and $b=1-f^{1/2}\ggg$ the result becomes
\be
\Fc(f<1/4)=\Fa\frac{f^{-1/2}}{\gamma_1K_0(s-1)} \left[ 1-g_2^{1-s} \right] \label{f1s3a01c1} \eqd
\ee
For $f>1/4$ we obtain for $a=f^{1/2}$ and $b=1-f^{1/2}\ggg$
\be
\Fc(f>1/4)=\Fa\frac{f^{-1/2}}{\gamma_1K_0(s-1)} \nonumber \\
\times \left[ (f^{-1/2}-1)^{s-1}-g_2^{1-s} \right] \label{f1s3a01c2} \eqc
\ee
which becomes negative for $f>1$. \\
Collecting terms, we obtain the total fluence for this case
\be
\Fc(1<s<3,\alpha_0\ll 1)=\Fa \left[ F_c^4(f<1/4)\right. \nonumber \\
+F_c^5(1/4<f<1)+F_c^6(1<f<1/4g_2^2) \nonumber \\
\left. +F_c^7(1/4g_2^2<f<g_2^2) \right] \label{f1s3a01sol} \eqc
\ee
with
\be
F_c^4 &=& \frac{f^{-1/2}}{\gamma_1K_0(s-1)} \left[ 1-g_2^{1-s} \right] \\
F_c^5 &=& \frac{f^{-1/2}}{\gamma_1K_0(s-1)} \left[ (f^{-1/2}-1)^{s-1}-g_2^{1-s} \right] \nonumber \\
& &+ \frac{f^{-1/2}\alpha_0^2}{2\gamma_1U_1\beta}\left[ 1-(f^{-1/2}-1)^{s-1} \right] \nonumber \\
&=& \frac{f^{-1/2}}{\gamma_1K_0(s-1)} \left[ 1-g_2^{1-s} \right] = F_c^4 \\
F_c^6 &=& \frac{f^{-1/2}}{\gamma_1(s-1)} \left\{ \frac{1-\left( 1-\left( \frac{f}{g_2^2} \right)^{1/2} \right)^{s-1}}{K_0+U_1g_2^{3-s}}\right. \nonumber \\
& & \left. + \frac{\left( 1-\left( \frac{f}{g_2^2} \right)^{1/2} \right)^{s-1}-\left( \frac{f}{g_2^2} \right)^{\frac{s-1}{2}}}{K_0} \right\} \\
F_c^7 &=& \frac{f^{-s/2}}{\gamma_1(K_0+U_1g_2^{3-s})(s-1)}\left[ 1-\left( \frac{f}{g_2^2} \right)^{\frac{s-1}{2}} \right] \eqd
\ee
In $F_c^5$ we inserted the definitions of $\alpha_0$ and $\beta$, which made it possible to find $F_c^5=F_c^4$. \\
The main results are similar to the steep injection. For $f<1$ the spectrum shows a power-law with a constant spectral index $\beta_1=1/2$, while for $f>1$ the spectrum steepens with $\Delta\beta_1=(s-1)/2$. We also find the cut-off at $f=g_2^2$.


\subsection{Numerical solution for a small injection parameter} \label{ssec:numsi}

Before we proceed with $\alpha_0\gg 1$, we want to compare the analytical result with a simple numerical integration, as we did for the linear and non-linear cooling, as well. For the plots we dropped the constant $\Fa$, which means that we normalized the fluence by $N\Fc=\Fc/\Fa$. \\
In the following plots we fix $\alpha_0$, and, thus, $q_0$ is not a free parameter any more, but becomes
\be
q_0=\frac{\alpha_0^2K_0(s-1)}{\gamma_1^{3-s}(1-g_2^{1-s})} \eqd
\ee
This confirms what we discussed in section \ref{ssec:teco}, that $q_0$ strongly influences which type of cooling occurs. 
\begin{figure}[ht]
	\centering
		\includegraphics[width=0.50\textwidth]{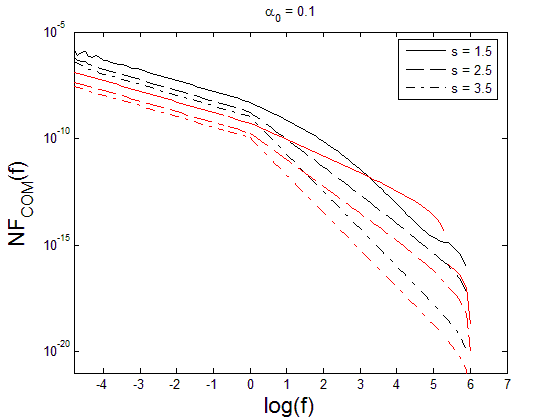}
	\caption{$NF_{COM}$ as a function of $f$ for three cases of $s$ (\textit{Full:} $s=1.5$, \textit{Dashed:} $s=2.5$, \textit{Dot-dashed:} $s=3.5$) and for $\alpha_0=0.1$, and $g_2=10^3$. The red lines indicate the analytical solution Eqs. (\ref{fs3a01sol}) and (\ref{f1s3a01sol}).}
	\label{fig:COMa01}
\end{figure}\\
\begin{figure}[ht]
	\centering
		\includegraphics[width=0.50\textwidth]{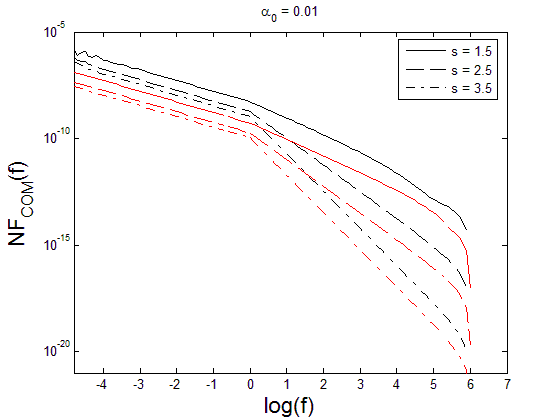}
	\caption{$NF_{COM}$ as a function of $f$ for three cases of $s$ (\textit{Full:} $s=1.5$, \textit{Dashed:} $s=2.5$, \textit{Dot-dashed:} $s=3.5$) and for $\alpha_0=0.01$, and $g_2=10^3$. The red lines indicate the analytical solution Eqs. (\ref{fs3a01sol}) and (\ref{f1s3a01sol}).}
	\label{fig:COMa001}
\end{figure}\\
The results shown in Figs. \ref{fig:COMa01} and \ref{fig:COMa001} do not match each other as well as they did in the linear and non-linear cooling cases. Although we inserted no offset, the analytical solution has a smaller magnitude than the numerical result. \\
For frequencies below unity one can see, however, that the behavior of the analytical and numerical solutions are the same. Both show a power-law with a spectral index of $\beta_1=1/2$. For frequencies above unity and for $s>2$ the results also agree rather well, and show the cut-off at $f=g_2^2$. \\ 
For $s<2$ the results do not match each other so well any more. The numerical result cannot be represented as a single power-law, but steepens gradually until it reaches a small plateau (which one could call a small pile-up) at about $f=1/4g_2^2$, and cuts off at $f=g_2^2$. This behavior is more obvious for $\alpha_0$ closer to unity, which could be a hint that the approximations are not valid enough for this case. Interestingly, for $s=1.5$ and $\alpha_0=0.1$ the analytical solution is not continuous at $f=1/4g_2^2$, which is a little bit strange, since it should be continuous as one can easily verify.


\subsection{Large spectral index, large injection parameter} \label{ssec:lsli}

We will continue our analytical calculation of the synchrotron fluence with the case of $\alpha_0\gg 1$, and begin with $s>3$. \\
For $0<\Psi<\Psi_2$ the calculation is the same as for $\alpha\ll 1$, which means that we can use the results obtained during the calculation of that case. Thus,
\be
\Fc(1/4<f<1)=\Fa \frac{f^{-1/2}}{\gamma_1(K_0+U_0)(s-1)} \nonumber \\
\times \left[ 1-\left( f^{-1/2}-1 \right)^{s-1} \right] \label{fs3a10a1} \eqc
\ee
and
\be
\Fc(1<f<g_2^2)=\Fa \frac{f^{-s/2}}{\gamma_1(K_0+U_0)(s-1)} \nonumber \\
\times \left[ 1-\left( \frac{f}{g_2^2} \right)^{\frac{s-1}{2}} \right] \label{fs3a10a2} \eqd
\ee
The next integration will be done for $\Psi_2\leq\Psi\leq\Psi_3$, where $\Psi_3=f^{1/2}\alpha_0$. Hence, the differential becomes $\td{\Psi}=\Psi^{-2}f^{3/2}\alpha_0^2\gamma_1K_0\td{y}$, and the fluence reads
\be
\Fc(f)=\Fa\frac{f^{-\frac{2+s}{2}}}{\alpha_0^2\gamma_1K_0}\intl_{\Psi_2}^{\Psi_3}\td{\Psi}\Psi^2(1-\Psi)^{s-2} \nonumber \\
\times \HSF{(1-f^{1/2}\ggg)-\Psi}\HSF{\Psi-(1-f^{1/2})} \nonumber \\
=\Fa\frac{f^{-\frac{2+s}{2}}}{\alpha_0^2\gamma_1K_0} \left[ -\frac{\Psi^2}{s-1}(1-\Psi)^{s-1} \right. \nonumber \\
\left. -\frac{2\Psi}{s^2-s}(1-\Psi)^s-\frac{2}{s^3-s}(1-\Psi)^{s+1} \right]_a^b \nonumber \\
\approx \Fa\frac{f^{-\frac{2+s}{2}}}{\alpha_0^2\gamma_1K_0} \left[ -\frac{\Psi^2}{s-1}(1-\Psi)^{s-1} \right]_a^b \label{fs3a10b} \eqd
\ee
In the last step we used the leading term approximation again, which we have already used during the calculations of the non-linear fluence. We infer from the limits and the Heaviside functions that the fluence of this part could contain three frequency regimes. \\
The first one is for $f<\alpha_0^{-2}$, and with $a=1-f^{1/2}$ and $b=f^{1/2}\alpha_0^2$ we obtain
\be
\Fc(f<\alpha_0^{-2}) = \Fa\frac{f^{-3/2}}{\alpha_0^2\gamma_1K_0(s-1)} \nonumber \\
\times \left[ (1-f^{1/2})^2-f\alpha_0^2(f^{-1/2}-\alpha_0)^{s-1} \right] \label{fs3a10b1} \eqd
\ee
Inspecting this solution a little further one finds that this does not contribute to the overall fluence for $f<(\alpha_0+1)^{-2}$. This means that this solution has only a very narrow regime of applicability. Therefore, we neglect it entirely. \\
The second frequency interval is $\alpha_0^{-2}<f<1/4$ with the limits $a=1-f^{1/2}$, and $b=1-f^{1/2}\ggg$. Hence,
\be
\Fc(\alpha_0^{-2}<f<1/4) = \Fa \frac{f^{-3/2}}{\alpha_0^2\gamma_1K_0(s-1)} \nonumber \\
\times \left[ \left( 1-f^{1/2} \right)^2-\left( 1-\left( \frac{f}{g_2^2} \right)^{1/2} \right)^2 g_2^{1-s} \right] \nonumber \\
\approx \Fa \frac{f^{-3/2}}{\alpha_0^2\gamma_1K_0(s-1)} \left[ 1-g_2^{1-s} \right] \label{fs3a10b2} \eqd
\ee
The last interval is for $f>1/4$, and with $a=f^{1/2}$ and $b=1-f^{1/2}\ggg$ we find
\be
\Fc(f>1/4) = \Fa \frac{f^{-\frac{2+s}{2}}}{\alpha_0^2\gamma_1K_0(s-1)} \nonumber \\
\times \left[ f \left( 1-f^{1/2} \right)^{s-1} - \left( 1- \left( \frac{f}{g_2^2} \right)^{1/2} \right)^2 f^{\frac{s-1}{2}} g_2^{1-s} \right] \nonumber \\
\approx \Fa \frac{f^{-s/2}}{\alpha_0^2\gamma_1K_0(s-1)} \left[ 1-f^{1/2} \right]^{s-1} \label{fs3a10b3}
\ee
This solution turns negative for $f>1$. \\
One integration remains, which is for $\Psi>\Psi_3$. The differential is substituted by $\td{\Psi}=f^{1/2}\gamma_1K_0\td{y}$. Thus,
\be
\Fc(f)=\Fa\frac{f^{-s/2}}{\gamma_1K_0}\intl_{\Psi_3}^{\infty}\td{\Psi}(1-\Psi)^{s-2} \nonumber \\
\times \HSF{(1-f^{1/2}\ggg)-\Psi}\HSF{\Psi-(1-f^{1/2})} \nonumber \\
=\Fa\frac{f^{-s/2}}{\gamma_1K_0(s-1)} \left[ -(1-\Psi) \right]_a^b \label{fs3a10c} \eqd
\ee
For $f<\alpha_0^{-2}$ the limits become $a=1-f^{1/2}$ and $b=1-f^{1/2}\ggg$, and, therefore,
\be
\Fc(f<\alpha_0^{-2}) = \Fa \frac{f^{-1/2}}{\gamma_1K_0(s-1)} \left[ 1-g_2^{1-s} \right] \label{fs3a10c1} \eqd
\ee
In the case $f>\alpha_0^{-2}$ $a=f^{1/2}\alpha_0$, and $b=1-f^{1/2}\ggg$ resulting in
\be
\Fc(f>\alpha_0^{-2}) = \Fa \frac{f^{-s/2}}{\gamma_1K_0(s-1)} \nonumber \\
\times \left[ \left( 1-f^{1/2}\alpha_0 \right)^{s-1} - f^{\frac{s-1}{2}}g_2^{1-s} \right] \label{fs3a10c2} \eqc
\ee
which is always negative, and does not contribute to the fluence. \\
Collecting terms the fluence for $s>3$ and $\alpha_0\gg 1$ becomes
\be
\Fc(s>3,\alpha_0\gg 1) = \Fa \left[ \Fs{1}(f<\alpha_0^{-2}) \right. \nonumber \\
+\Fs{2}(\alpha_0^{-2}<f<1/4)+\Fs{3}(1/4<f<1) \nonumber \\
\left. +\Fs{4}(1<f<g_2^2) \right] \label{fs3a10sol} \eqc
\ee
with
\be
\Fs{1} &=& \frac{f^{-1/2}}{\gamma_1K_0(s-1)} \left[ 1-g_2^{1-s} \right]  \\
\Fs{2} &=& \frac{f^{-3/2}}{\alpha_0^2\gamma_1K_0(s-1)} \left[ 1-g_2^{1-s} \right] \\
\Fs{3} &=& \frac{f^{-1/2}}{\gamma_1(s-1)}\left[ \frac{K_0+U_0\left( f^{-1/2}-1 \right)^{s-1}}{(K_0+U_0)K_0} \right] \\
\Fs{4} &=& \frac{f^{-s/2}}{\gamma_1(K_0+U_0)(s-1)} \left[ 1-\left( \frac{f}{g_2^2} \right)^{\frac{s-1}{2}} \right] \eqd
\ee
What we have found is a spectrum that is represented by mainly three power-laws, instead of two as in the previous (especially the linear and non-linear) cases. For very low frequency the spectral index is $\beta_1=1/2$, identically to the linear cooling as expected. For intermediate frequency up to $f\approx 1$ the spectral index becomes $\beta_2=3/2$, which is also expected since this is identical to the non-linearly cooled case. In the region around unity the spectrum changes its spectral index according to $\Delta\beta_2=(s-3)/2$. The spectrum is cut off at $f=g_2^2$.


\subsection{Small spectral index, large injection parameter} \label{ssec:ssli}

Finally, we deal with the case $1<s<3$ and $\alpha_0\gg 1$. \\
For $\Psi<\Psis_1$ we can use the result obtained in section \ref{ssec:sssi}, yielding
\be
\Fc(1<f<1/4g_2^2) = \Fa\frac{f^{-s/2}}{\gamma_1(K_o+U_1g_2^{3-s})(s-1)} \nonumber \\
\times \left[ 1-\left( 1-\left( \frac{f}{g_2^2} \right)^{1/2} \right)^{s-1} \right] \label{f1s3a10a1} \eqc
\ee
and
\be
\Fc(1/4g_2^2<f<g_2^2) = \Fa \frac{f^{-s/2}}{\gamma_1(K_o+U_1g_2^{3-s})(s-1)} \nonumber \\
\times \left[ 1-\left( \frac{f}{g_2^2} \right)^{\frac{s-1}{2}} \right] \label{f1s3a10a2} \eqd
\ee
For $\Psis_1<\Psi<\Psis_3$ with $\Psis_3\approx f^{1/2}$, we substitute the derivative with 
\be
\td{\Psi}=\Psi{3-s}f^{\frac{s-2}{2}}\left( \frac{s-3}{4-s} \right)\left( \frac{2\gamma_1U_1(4-s)}{s-1} \right)^{\frac{s-2}{4-s}} \td{y} \eqc
\ee
resulting in
\be
\Fc(f)=\Fa\frac{f^{\frac{3-s}{2}}(4-s)(s-1)^{\frac{s-2}{4-s}}}{(s-3)(2\gamma_1U_1(4-s))^{\frac{s-2}{4-s}}} \intl_{\Psis_1}^{\Psis_3}\td{\Psi} \Psi^{s-3} \nonumber \\
\times (1-\Psi)^{s-2} \HSF{(1-f^{1/2}\ggg)-\Psi} \nonumber \\
\times \HSF{\Psi-(1-f^{1/2})} \label{f1s3a10b} \eqd
\ee
The integral can be expressed in terms of the hypergeometric function, but, unfortunately, one cannot obtain an analytical form. Hence, we have to terminate further discussions of this case. \\
The next time step is $\Psis_3\leq\Psi\leq\Psis_4$, where $\Psi_4\approx f^{1/2}\alpha_0$. Thus, the differential becomes $\td{\Psi}=\Psi^{-2}f^{3/2}\alpha_0^2\gamma_1K_0\td{y}$, and the fluence reads
\be
\Fc(f)=\Fa\frac{f^{-\frac{2+s}{2}}}{\alpha_0^2\gamma_1K_0}\intl_{\Psis_3}^{\Psis_4}\td{\Psi}\Psi^2(1-\Psi)^{s-2} \nonumber \\
\times \HSF{(1-f^{1/2}\ggg)-\Psi}\HSF{\Psi-(1-f^{1/2})} \nonumber \\
=\Fa\frac{f^{-\frac{2+s}{2}}}{\alpha_0^2\gamma_1K_0} \left[ -\frac{\Psi^2}{s-1}(1-\Psi)^{s-1} \right. \nonumber \\
\left. -\frac{2\Psi}{s^2-s}(1-\Psi)^s-\frac{2}{s^3-s}(1-\Psi)^{s+1} \right]_a^b \nonumber \\
\approx \Fa\frac{f^{-\frac{2+s}{2}}}{\alpha_0^2\gamma_1K_0} \left[ -\frac{\Psi^2}{s-1}(1-\Psi)^{s-1} \right]_a^b \label{f1s3a10c} \eqd
\ee
This is identical to Eq. (\ref{fs3a10b}) and we used the same approximations, again, as in that calculation. Therefore, the final results will be the same, as well, and we can just copy them from Eqs. (\ref{fs3a10b2}) and (\ref{fs3a10b3}), neglecting Eq. (\ref{fs3a10b1}) as before:
\be
\Fc(\alpha_0^{-2}<f<1/4) = \Fa \frac{f^{-3/2}}{\alpha_0^2\gamma_1K_0(s-1)} \nonumber \\
\times \left[ \left( 1-f^{1/2} \right)^2-\left( 1-\left( \frac{f}{g_2^2} \right)^{1/2} \right)^2 g_2^{1-s} \right] \nonumber \\
\approx \Fa \frac{f^{-3/2}}{\alpha_0^2\gamma_1K_0(s-1)} \left[ 1-g_2^{1-s} \right] \label{f1s3a10c2} \eqc
\ee
and
\be
\Fc(f>1/4) = \Fa \frac{f^{-\frac{2+s}{2}}}{\alpha_0^2\gamma_1K_0(s-1)} \nonumber \\
\times \left[ f \left( 1-f^{1/2} \right)^{s-1} - \left( 1- \left( \frac{f}{g_2^2} \right)^{1/2} \right)^2 f^{\frac{s-1}{2}} g_2^{1-s} \right] \nonumber \\
\approx \Fa \frac{f^{-s/2}}{\alpha_0^2\gamma_1K_0(s-1)} \left[ 1-f^{1/2} \right]^{s-1} \label{f1s3a10c3} \eqd
\ee
This solution turns negative for $f>1$. \\
The last integration is for $\Psi>\Psis_4$ with the differential $\td{\Psi}=f^{1/2}\gamma_1K_0\td{y}$. Thus,
\be
\Fc(f)=\Fa\frac{f^{-s/2}}{\gamma_1K_0}\intl_{\Psis_4}^{\infty}\td{\Psi}(1-\Psi)^{s-2} \nonumber \\
\times \HSF{(1-f^{1/2}\ggg)-\Psi}\HSF{\Psi-(1-f^{1/2})} \nonumber \\
=\Fa\frac{f^{-s/2}}{\gamma_1K_0(s-1)} \left[ -(1-\Psi) \right]_a^b \label{f1s3a10d} \eqd
\ee
We can use the solution (\ref{fs3a10c1}) from the case $s>3$, yielding
\be
\Fc(f<\alpha_0^{-2}) = \Fa \frac{f^{-1/2}}{\gamma_1K_0(s-1)} \left[ 1-g_2^{1-s} \right] \label{f1s3a10d1} \eqd
\ee
Since the result of Eq. (\ref{fs3a10c2}) is always negative, we neglect it here, too. \\
Collecting terms we find
\be
\Fc(1<s<3,\alpha_0\gg 1) = \Fa \left[ \Fs{5}(f<\alpha_0^{-2}) \right. \nonumber \\
\left. +\Fs{6}(\alpha_0^{-2}<f<1/4)+\Fs{7}(1/4g_2^2<f<g_2^2) \right] \label{f1s3a10sol} \eqc
\ee
with
\be
\Fs{5} &=& \frac{f^{-1/2}}{\gamma_1K_0(s-1)} \left[ 1-g_2^{1-s} \right]  \\
\Fs{6} &=& \frac{f^{-3/2}}{\alpha_0^2\gamma_1K_0(s-1)} \left[ 1-g_2^{1-s} \right] \\
\Fs{7} &=& \frac{f^{-s/2}}{\gamma_1(K_0+U_1g_2^{3-s})(s-1)} \left[ 1-\left( \frac{f}{g_2^2} \right)^{\frac{s-1}{2}} \right] \eqd
\ee
Since the frequency regime $1/4<f<1/4g_2^2$ is influenced by the solution of the part $\Psis_1<\Psi<\Psis_3$, for which we could not find an analytical expression, we neglected the solutions in the overall result of the fluence (Eq. (\ref{f1s3a10sol})). \\
Nonetheless, the obtained result is noteworthy. As a matter of fact, it is remarkably similar to the case with $s>3$ and $\alpha_0\gg 1$. The fluence of Eq. (\ref{f1s3a10sol}) is also represented by a broken power-law, with two different but constant spectral indices $\beta_1=1/2$ and $\beta_2=3/2$ for frequencies below unity, indicating that the non-linear cooling is replaced by the linear cooling for very late times. The spectrum is cut off at $f=g_2^2$, similar to every calculated case before.


\subsection{Numerical solution for a large injection parameter} \label{ssec:numli}

As we did for $\alpha\ll 1$, we want to compare our analytical result for $\alpha_0\gg 1$ with a simple numerical calculation. For the plots we dropped the constant $\Fa$, which means that we normalized the fluence by $N\Fc=\Fc/\Fa$.
\begin{figure}[ht]
	\centering
		\includegraphics[width=0.50\textwidth]{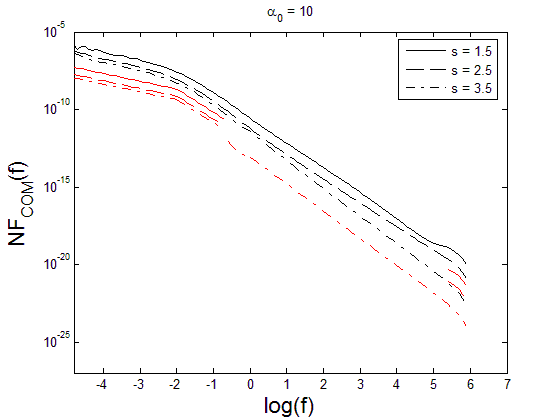}
	\caption{$NF_{COM}$ as a function of $f$ for three cases of $s$ (\textit{Full:} $s=1.5$, \textit{Dashed:} $s=2.5$, \textit{Dot-dashed:} $s=3.5$) and for $\alpha_0=10$, and $g_2=10^3$. The red lines indicate the analytical solution Eqs. (\ref{fs3a10sol}) and (\ref{f1s3a10sol}), but with an offset of $10^{-4}$.}
	\label{fig:COMa10}
\end{figure}\\
\begin{figure}[ht]
	\centering
		\includegraphics[width=0.50\textwidth]{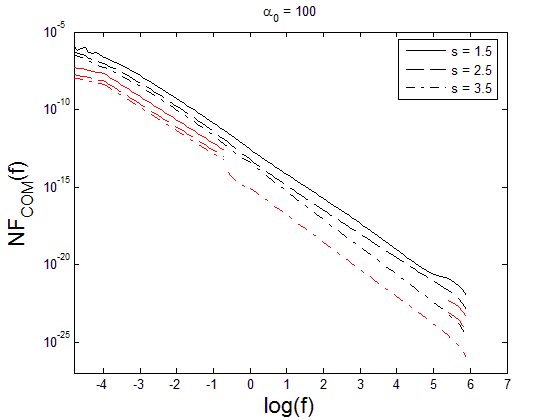}
	\caption{$NF_{COM}$ as a function of $f$ for three cases of $s$ (\textit{Full:} $s=1.5$, \textit{Dashed:} $s=2.5$, \textit{Dot-dashed:} $s=3.5$) and for $\alpha_0=100$, and $g_2=10^3$. The red lines indicate the analytical solution Eqs. (\ref{fs3a10sol}) and (\ref{f1s3a10sol}), but with an offset of $10^{-4}$.}
	\label{fig:COMa100}
\end{figure}\\
The results shown in Figs. \ref{fig:COMa10} and \ref{fig:COMa100} match each other rather well for $f<1$. The point $f=\alpha_0^{-2}$, where the power-law breaks from the linear to the non-linear cooling, is visible in the numerical and the analytical plots. However, in the numerical plot the spectral index for $f<\alpha_0^{-2}$ is a little higher than $1/2$. \\
For $f>1$ and $s>2$ the numerical plots seem to follow an $f^{-s/2}$ dependence, and one should note especially that for $s=2.5$ the plot is not as steep for $f>1$ as for $f<1$. This is expected, when we assume that the spectral index changes at $f=1$ like $\Delta\beta_2=(s-3)/2$. \\
For $s<2$ the curve shows a similar behavior as in the case $\alpha_0\ll 1$, steepening gradually for $f>1$, although the steepening is not as obvious as in the other case (cf. section \ref{ssec:numsi}). One should also note the small pile-up for $s=1.5$ shortly before the cut-off, which occurs at $f=g_2^2$ in all cases. \\

\nid At last we should note that the results of section \ref{sec:sfcom} match those of SBM, at least for $f<1$. This is reasonable, since the power-law electron distribution quenches rather rapidly until it becomes a $\delta$-function, which was used by SBM.


\section{Discussion and conclusions} \label{sec:dac}

In this paper we have made further progress in the work on nonlinear cooling of highly relativistic electrons, as it could take place in jets of active galactic nuclei. We used the nonlinear approach of RS, who introduced a cooling scenario based upon the synchrotron self-Compton effect. We expanded that paper by using a power-law for the initial electron distribution instead of a $\delta$-function. \\
We treated both the linear and the nonlinear cooling case, and calculated the respective electron number densities in section \ref{sec:end}, which differed by a new time variable that entered the nonlinear solution. We showed that the nonlinear solution cools much faster for harder spectra than for softer ones. The linear solution also cooled rather rapidly. We discussed in section \ref{ssec:teco} that it depends critically on the choice of some parameters, which cooling acts quicker. This implies that the approach treating the cooling scenarios separately is not necessarily fulfilled, because at one moment in time the linear cooling will always dominate over the nonlinear case. \\
Therefore, we applied the electron distribution to a cooling scenario, which combines the linear and non-linear approach in one term. This was also already calculated for a $\delta$-distribution of electrons by SBM. We calculated the electron number density in section \ref{sec:ccend}, and discussed that depending on the injection parameter $\alpha_0$ the electron number density showed characteristics of linear or non-linear cooling. However, even if the cooling is non-linear in the beginning, it becomes linear at later times, just as we expected. \\
We calculated the total synchrotron fluence for all cases and began with the linear and non-linear cases in sections \ref{sec:lsf} and \ref{sec:sfs}, respectively. In both cases the low-frequency regime of the fluence spectrum ($f<1$) does not depend on the injection spectral index $s$ of the electrons. The fluence spectral index is $\alpha_{SYN}=1/2$ for linear, and $\alpha_{SSC}=3/2$ for nonlinear cooling, respectively. These constant spectral indices match those of the $\delta$-function approach for the electron number density used by RS. This is reasonable, since we showed in section \ref{ssec:teco} that the electron distribution becomes a $\delta$-function for late times, when the electrons have cooled significantly and the radiated synchrotron photon energy is low. \\
For linear as well as for nonlinear cooling the spectra are cut off at $f=g_2^2$. In the region $1<f<g_2^2$ the spectral index is $s/2$ with the exception of the nonlinear case with small spectral indices, where the spectral index remains basically unchanged from the low-energetic part. \\
Section \ref{sec:sfcom} was devoted to the calculation of the total synchrotron fluence of combined cooled electrons. The results are in some ways similar to the cases before. For $\alpha_0\ll 1$, which represents some kind of a linear solution, the fluence shows linear characteristics. Especially, below $f=1$ the spectral index is $\beta_1=\alpha_{SYN}=1/2$. For $\alpha_0\gg 1$, the fluence below $f=1$ exhibits a power-law with a spectral index $\beta_2=\alpha_{SSC}=3/2$, just as one would expect from a non-linear cooling. However, since the non-linear cooling becomes linear at later times, for frequencies below $f=\alpha_0^{-2}\ll 1$ the spectral index becomes $\beta_1=\alpha_{SYN}=1/2$, again. \\
Our results prove that the spectral behavior of the total synchrotron fluences in the non-linear SSC and combined synchrotron-SSC cooling cases is practically independent from the functional form of the energy injection spectrum. The earlier predictions of SBM, based on a mono-energetic injection distribution functions of electrons, on the synchrotron fluence behavior therefore hold in this case. 


%
\begin{acknowledgements} 
We acknowledge support from the German Ministry for Education and Research (BMBF) through Verbundforschung Astroteilchenphysik grant 05A08PC1 and 
the Deutsche Forschungsgemeinschaft through grants Schl 201/20-1 and Schl 201/23-1.  
\end{acknowledgements}
%


\appendix
\section{Calculation of the non-linear electron number density} \label{sec:append}

Inserting the nonlinear cooling rate Eq. (\ref{ssccool}) and the injection rate into Eq. (\ref{eldispde}) gives us the differential equation for the SSC cooled electron number density $n_{SSC}$ (we drop the subscript in the following):
\be
\frac{\pd{n(\gamma,t)}}{\pd{t}}-\frac{\pd}{\pd{\gamma}}\left( A_0\gamma^2\left[\intl_{0}^{\infty}\td{\gamma}\gamma^2 n(\gamma,t)\right] n(\gamma,t) \right) \nonumber \\
= q_0\gamma^{-s}\HSF{\gamma-\gamma_1}\HSF{\gamma_2-\gamma}\DELF{t} \label{nleldispde1} \eqd
\ee 
Multiplying the equation with $\gamma^2/A_0$ we obtain with the definitions $y=A_0t$ and $S=\gamma^2n$
\be
\frac{\pd{S}}{\pd{y}}-\left[ \intl_0^{\infty}\td{\gamma}S \right]\gamma^2\frac{\pd{S}}{\pd{\gamma}} \nonumber \\
=q_0\gamma^{2-s} \DELF{y} \HSF{\gamma_2-\gamma}\HSF{\gamma-\gamma_1} \label{nleldispde2} \eqd
\ee
We yield with $\xi=\gamma^{-1}$
\be
\frac{\pd{S}}{\pd{y}}+\left[ \intl_0^{\infty}\td{\xi} \frac{S}{\xi^2} \right] \frac{\pd{S}}{\pd{\xi}} \nonumber \\
=q_0\xi^{s-2}\DELF{y}\HSF{\xi- \xi_2}\HSF{\xi_1-\xi} \label{nleldispde3} \eqd
\ee
If we define the implicit time variable $T$ through
\be
\frac{\td{T}}{\td{y}}=\intl_{0}^{\infty}\td{\xi}\frac{S}{\xi^2} \label{impltimevariabapp} \eqc
\ee
the differential equation becomes
\be
\frac{\pd{S}}{\pd{y}}+\frac{\td{T}}{\td{y}} \frac{\pd{S}}{\pd{\xi}}=q_0\xi^{s-2}\DELF{y}\HSF{\xi- \xi_2}\HSF{\xi_1-\xi} \label{nleldispde4} \eqd
\ee
Formally multiplying this equation with $\td{y}/\td{T}$ results in
\be
\frac{\pd{S}}{\pd{T}}+ \frac{\pd{S}}{\pd{\xi}}=q_0\xi^{s-2}\DELF{T}\HSF{\xi- \xi_2}\HSF{\xi_1-\xi} \label{nleldispde6} \eqd
\ee
This differential equation for the electron number density can be solved with the method of characteristics. Thus , we obtain
\be
\frac{\td{S}}{\td{T}}=q_0 (T+\xi_0)^{s-2}\DELF{T}\HSF{T+\xi_0-\xi_2} \nonumber \\
\times \HSF{\xi_1-T-\xi_0} \label{methodcharact3} \eqc
\ee
where $\xi_0=\xi-T$ is a constant of integration. Eq. (\ref{methodcharact3}) can be easily integrated with respect to $T$, which results in
\be
S = q_0 (\xi-T)^{s-2}\HSF{\xi-T-\xi_2}\HSF{\xi_1-\xi+T}\HSF{T} \nonumber \\
+S_h \label{solmethodcharact1} \eqd
\ee
We now require that $S(\xi=0,T)=0$, which means
\be
S_h(-T) = -q_0 (-T)^{s-2}\HSF{-T-\xi_2}\HSF{\xi_1+T}\HSF{T} \eqd
\ee
Collecting terms, we find $S$ to be
\be
S(\xi,T) = q_0 (\xi-T)^{s-2}\HSF{\xi-T-\xi_2}\HSF{\xi_1-\xi+T} \nonumber \\
\times \left( \HSF{T}-\HSF{T-\xi} \right) \label{solmethodcharact2} \eqd
\ee
Since our flare begins at $T=0$, we are not interested in events that take place before that moment. Hence, we find the electron number density:
\be
n(\gamma,T\geq 0) = \frac{S}{\gamma^2} \nonumber \\
= q_0 \gamma^{-s} \left( 1-\gamma T \right)^{s-2}\HSF{\frac{1}{T} -\gamma} \nonumber \\
\times \HSF{\frac{\gamma_2}{1+\gamma_2 T} -\gamma}\HSF{\gamma- \frac{\gamma_1}{1+\gamma_1 T}}  \label{nleldis1} \eqd
\ee
We see that we have two Heaviside functions defining upper limits for $\gamma$. It is an easy task to compare them, and to find out which one is lower than the other one. Having done so, we find the solution for the nonlinearly cooled electron number density to be
\be
n(\gamma,T)=q_0\gamma^{-s}\left( 1-\gamma T \right)^{s-2}\HSF{\frac{\gamma_2}{1+\gamma_2 T} -\gamma} \nonumber \\
\times \HSF{\gamma- \frac{\gamma_1}{1+\gamma_1 T}}  \label{nleldisAPP} \eqc
\ee
yielding Eq. (\ref{nleldis}).


\section{Derivation of U} \label{sec:appu}

The time variable $T$ has been defined through Eq. (\ref{impltimevariab}):
\be
U(T)\equiv \frac{\td{T}}{\td{y}}=\intl_{0}^{\infty}\td{\xi}\frac{S}{\xi^2} \eqd
\ee
Inserting Eq. (\ref{solmethodcharact2}), we gain
\be
U(T) = \intl_{0}^{\infty}\td{\xi}\xi^{-2}q_0(\xi-T)^{s-2}\HSF{\xi-T-\xi_2} \nonumber \\
\times \HSF{\xi_1+T-\xi}\HSF{T}  \label{impltimecalc1} \eqd 
\ee
As stated before, we are only interested in solutions for $T\geq 0$. Hence, we can neglect the third Heaviside function, which results in
\be
U(T) = q_0\intl_{T+\xi_2}^{T+\xi_1}\td{\xi}\xi{-2}(\xi-T)^{s-2} \label{impltimecalc2} \eqd
\ee
A first substitution $w=\xi-T$ yields
\be
U(T)=q_0\intl_{\xi_2}^{\xi_1}\td{w}(w+T)^{-s}w^{s-2} \label{impltimecalc3} \eqc
\ee
while a second substitution $w=Tv$ gives
\be
U(T) = q_0\intl_{\xi_2/T}^{\xi_1/T}\td{v}TT^{-2}(1+v)^{-2}T^{s-2}v^{s-2} \nonumber \\
= q_0 T^{s-3}\intl_{(\gamma_2 T)^{-1}}^{(\gamma_1 T)^{-1}} \td{v}\frac{v^{s-2}}{(1+v)^2} \label{impltimecalc4} \eqc
\ee
where we re-substituted $\xi_i=\gamma_i^{-1}$ in the last step. \\
The purpose of the next two substitutions is to get rid of the time variable in the limits of the integral. In order to achieve this we first set $x=\gamma_1 T$ and introduce $g_2=\gamma_2/\gamma_1$, which yields
\be
U(x) = q_0\gamma_{1}^{3-s}x^{s-3}\intl_{(g_2 x)^{-1}}^{x^{-1}}\td{v}\frac{v^{s-2}}{(1+v)^2} \nonumber \\
= q_0\gamma_1^{3-s}\intl_{(g_2 x)^{-1}}^{x^{-1}} \td{v} \frac{x^{-1}(xv)^{s-2}}{(1+v)^2} \label{impltimecalc5} \eqd
\ee
Now, we use $u=vx$, resulting in
\be
U(x) = q_0\gamma_1^{3-s}\intl_{g_2^{-1}}^{1}\td{u}\frac{x^{-2}u^{s-2}}{(1+\frac{u}{x})^2} \nonumber \\
= q_0\gamma_1^{3-s}\intl_{g_2^{-1}}^{1}\td{u}\frac{u^{s-2}}{(x+u)^2} \label{impltimecalc6} \eqd
\ee
This integral can be expressed in terms of the hypergeometric function, but that would not yield an analytical form. Nonetheless, one can obtain an approximate solution in the regimes $0\leq x\leq g_2^{-1}$ (small $x$), and $x\geq 1$ (large $x$). An analytical continuation serves as a solution for the intermediate regime. For small $x$ the integral can be written as
\be
U_1 = q_0\gamma_1^{3-s}\intl_{g_2^{-1}}^{1}\td{u}\frac{u^{s-2}}{(x+u)^2} \nonumber \\
\approx q_0\gamma_1^{3-s}\intl_{g_2^{-1}}^{1}\td{u}u^{s-4} \nonumber \\
= \frac{q_0\gamma_1^{3-s}}{s-3}\left[ 1-g_2^{3-s} \right] \label{impltimesolsmall} \eqd
\ee
Similarly, we achieve for large $x$
\be
U_3 = q_0\gamma_1^{3-s}\intl_{g_2^{-1}}^{1}\td{u}\frac{u^{s-2}}{(x+u)^2} \nonumber \\
\approx q_0\gamma_1^{3-s}\intl_{g_2^{-1}}^{1}\td{u}\frac{u^{s-2}}{x^2} \nonumber \\
= \frac{q_0\gamma_1^{3-s}}{(s-1)x^2} \left[ 1-g_2^{1-s} \right] \label{impltimesollarge} \eqd
\ee
The requirement for the solution of the intermediate regime is that it must be continuous, meaning $U_1(g_2^{-1})=U_2(g_2^{-1})$, and $U_2(1)=U_3(1)$. In order to accomplish such a behavior, we can first assume a proper solution $U_2$ with some unspecified constants, and then try to fit it to the boundary conditions. A good ansatz is
\be
U_2 = \frac{q_0\gamma_1^{3-s}}{s-3}\left[ a'-b'x^{s-3}-c'\frac{g_2^{1-s}}{x^2} \right] \label{impltimesolinte1} \eqd
\ee
Matching the solution with the boundary conditions, yields the values of the constants $a'$, $b'$, and $c'$:
\be
a' = 1 \nonumber \\
b' = \frac{2}{s-1} \nonumber \\
c' = \frac{s-3}{s-1} \nonumber \eqd
\ee
Since we had only two equations for three parameters, we chose $a'=1$. \\
Thus, the obtained solution for the intermediate $x$-range is
\be
U_2 = \frac{q_0\gamma_1^{3-s}}{s-3}\left[ 1-\frac{2x^{s-3}}{s-1}-\frac{s-3}{s-1}\frac{g_2^{1-s}}{x^2} \right] \label{impltimesolinte} \eqd
\ee
Before we summarize the results, we need to say a few words about the spectral index $s$. We already stated that it must be greater than $1$. But according to our results above, we also find that $s\neq 3$. Thus, we have two different cases to consider: $s>3$, and $1<s<3$. \\
Collecting terms, we find for $s>3$
\be
U(x) = \nonumber \\
\left\{ \begin{array}{ll} 
U_0 \left[ 1-g_2^{3-s} \right] & \eqc\, 0\leq x\leq g_2^{-1} \\ 
U_0 \left[ 1-\frac{2x^{s-3}}{s-1}-\frac{s-3}{s-1}\frac{g_2^{1-s}}{x^2} \right] & \eqc\, g_2^{-1}\leq x\leq 1 \\
U_0 \frac{s-3}{(s-1)x^2}\left[ 1-g_2^{1-s} \right] & \eqc\, x\geq 1  
\end{array} \right. \label{impltimesol1app} \eqc
\ee
and for $1<s<3$
\be
U(x) = \nonumber \\
\left\{ \begin{array}{ll} 
U_1 \left[ g_2^{3-s}-1 \right] & \eqc\, 0\leq x\leq g_2^{-1} \\ 
U_1 \left[ \frac{2x^{s-3}}{s-1}-1-\frac{s-3}{s-1}\frac{g_2^{1-s}}{x^2} \right] & \eqc\, g_2^{-1}\leq x\leq 1 \\
U_1 \frac{3-s}{(s-1)x^2}\left[ 1-g_2^{1-s} \right] & \eqc\, x\geq 1  
\end{array} \right. \label{impltimesol2app} \eqc
\ee
with $U_0=\frac{q_0\gamma_1^{3-s}}{s-3}$, and $U_1=\frac{q_0\gamma_1^{3-s}}{3-s}$. 


\section{The non-linear time variable x} \label{sec:appx}

Since $U(x)=\frac{\td{x}}{\gamma_1\td{y}}$, we can separate the variables obtaining $\td{y}=\frac{\td{x}}{\gamma_1 U(x)}$. This can be integrated quite easily except for both intermediate cases. However, we can find approximative solutions by using the same approximations of these cases we used already during the calculation of the synchrotron spectra. \\
Beginning with the case $s>3$ we yield
\be
y(0\leq x\leq \ggg)=\int \frac{\td{x}}{\gamma_1U_0(1-g_2^{3-s})} \nonumber \\
=\frac{1}{\gamma_1U_0(1-g_2^{3-s})}x+c_1 \label{x1solapp} \eqd
\ee
We require $y(x=0)=0$, which means $c_1=0$. The intermediate range becomes
\be
y(\ggg\leq x\leq 1)=\int \frac{\td{x}}{\gamma_1U_0\left( 1-\frac{2x^{s-3}}{s-1}-\frac{s-3}{s-1}\frac{g_2^{1-s}}{x^2} \right)} \nonumber \\
\approx \frac{1}{\gamma_1U_0}\int\td{x}=\frac{1}{\gamma_1U_0}x+c_2 \label{x2solapp} \eqd
\ee
\begin{figure}[ht]
	\centering
		\includegraphics[width=0.50\textwidth]{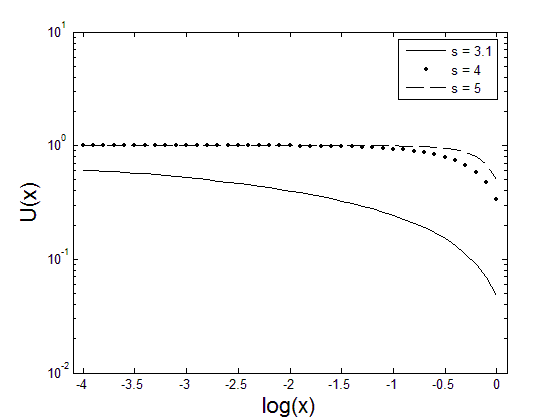}
	\caption{The denominator $U(x)$ of the intermediate time regime as a function of $x$ for three cases of $s$.}
	\label{fig:Uintes3a}
\end{figure}\\
We approximated the denominator with the leading term. The validity of this approximation for most parts of the $x$-range can be seen in Fig. \ref{fig:Uintes3a}. It becomes less valid for $s\rightarrow 3$. \\
For the third case we find
\be
y(x\geq 1)=\int \frac{\td{x}}{\gamma_1U_0\frac{s-3}{(s-1)x^2}(1-g_2^{1-s})} \nonumber \\
=\frac{s-1}{3\gamma_1U_0(s-3)(1-g_2^{1-s})}x^3+c_3 \label{x3solapp} \eqd
\ee
These equations can be inverted simply, yielding $x(y)$. As for $U(x)$, we require $x$ to be continuous at the points $y_1=y(x=\ggg)$ and $y_2=y(x=1)$. Matching the solutions at these points we find the values 
\be
y_1 &=& \frac{1}{\gamma_1 U_0} \frac{\ggg}{1-g_2^{3-s}} \\
y_2 &=& \frac{1}{\gamma_1 U_0} \left[ 1+\frac{g_2^{2-s}}{1-g_2^{3-s}} \right] \\
c_2 &=& \frac{1}{\gamma_1 U_0} \frac{g_2^{2-s}}{1-g_2^{3-s}} \\
c_3 &=& \frac{1}{\gamma_1 U_0} \left[ 1+\frac{g_2^{2-s}}{1-g_2^{3-s}}-\frac{s-1}{3(3-s)(1-g_2^{1-s})} \right] \eqd
\ee
For the case $1<s<3$ we find similarly
\be
y(0\leq x\leq \ggg)=\int \frac{\td{x}}{\gamma_1U_1(g_2^{3-s}-1)} \nonumber \\
=\frac{1}{\gamma_1U_1(g_2^{3-s}-1)}x+c_4 \label{x4solapp} \eqd
\ee
We require $y(x=0)=0$ again, which means $c_4=0$. The intermediate range becomes
\be
y(\ggg\leq x\leq 1)=\int \frac{\td{x}}{\gamma_1U_1\left( \frac{2x^{s-3}}{s-1}-1-\frac{s-3}{s-1}\frac{g_2^{1-s}}{x^2} \right)} \nonumber \\
\approx \frac{1}{\gamma_1U_0}\int\frac{\td{x}}{\frac{2x^{s-3}}{s-1}}=\frac{s-1}{2\gamma_1U_1(4-s)}x^{4-s}+c_5 \label{x5solapp} \eqd
\ee
\begin{figure}[ht]
	\centering
		\includegraphics[width=0.50\textwidth]{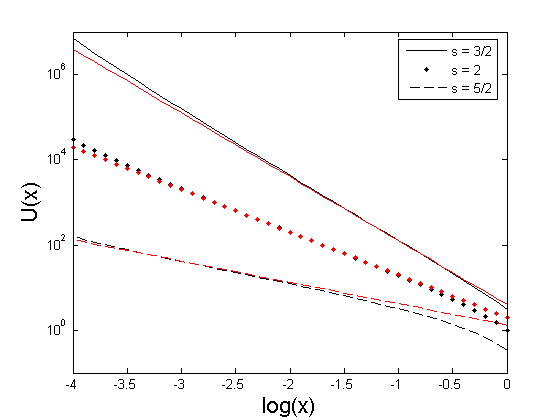}
	\caption{The denominator $U(x)$ (black) and its leading term (red) of the intermediate time regime as a function of $x$ for three cases of $s$.}
	\label{fig:Uinte1s3a}
\end{figure}\\
We approximated the denominator with the leading term. The validity of this approximation for most parts of the $x$-range can be seen in Fig. \ref{fig:Uinte1s3a}. The largest errors occur for small values of $x$. \\
The last case yields
\be
y(x\geq 1)=\int \frac{\td{x}}{\gamma_1U_1\frac{3-s}{(s-1)x^2}(1-g_2^{1-s})} \nonumber \\
=\frac{s-1}{3\gamma_1U_1(3-s)(1-g_2^{1-s})}x^3+c_6 \label{x6solapp} \eqd
\ee
As before, a simple inversion leads to $x(y)$, while the requirement that the solution should be continuous at the points $y_3=y(x=\ggg)$ and $y_4=y(x=1)$ gives the values
\be
y_3 &=& \frac{1}{\gamma_1 U_1} \frac{\ggg}{g_2^{3-s}-1} \\
y_4 &=& \frac{1}{\gamma_1 U_1} \left[ \frac{s-1}{2(4-s)}(1-g_2^{s-4})+\frac{\ggg}{g_2^{3-s}-1} \right] \\
c_5 &=& \frac{1}{\gamma_1 U_1} \left[ \frac{\ggg}{g_2^{3-s}-1}-\frac{s-1}{2(4-s)g_2^{4-s}} \right] \\
c_6 &=& \frac{1}{\gamma_1 U_1} \left[ \frac{s-1}{2(4-s)}(1-g_2^{s-4})+\frac{\ggg}{g_2^{3-s}-1} \right. \nonumber \\
		& & \left. -\frac{s-1}{3(3-s)(1-g_2^{1-s})} \right] \eqd
\ee


\section{The implicit time variable $\xc$ of the combined cooling} \label{sec:appxc}

The differential equation (\ref{eldispde}) may look a little bit more complicated with the combined cooling term (\ref{comcool}). However, the solution can be obtained with the methods outlined in appendix \ref{sec:append}, yielding solution (\ref{comeldis}). The important difference is the definition of the implicit time variable, which has to be chosen as
\be
\Us=\frac{\td{\tau}}{\td{y}}=K_0+\intl_{0}^{\infty}\td{\gamma}\gamma^{2}n_{COM}(\gamma,\tau) \label{app:implus1} \eqd
\ee
Using the definition of $U$ from appendix \ref{sec:appu}, this can be written as
\be
\Us=K_0+U \label{app:implus2} \eqd
\ee
Thus, we can use the previous results to obtain for $s>3$
\be
\Us(\xc,s>3) = \nonumber \\
\left\{ \begin{array}{ll} 
K_0+U_0 \left[ 1-g_2^{3-s} \right] & \eqc\, 0\leq \xc\leq g_2^{-1} \\ 
K_0+U_0  & \eqc\, g_2^{-1}\leq \xc\leq 1 \\
K_0+U_0 \frac{s-3}{(s-1)\xc^2}\left[ 1-g_2^{1-s} \right] & \eqc\, \xc\geq 1  
\end{array} \right. \label{app:implussol1} \eqc
\ee
and for $1<s<3$
\be
\Us(\xc,1<s<3) = \nonumber \\
\left\{ \begin{array}{ll} 
K_0+U_1 \left[ g_2^{3-s}-1 \right] & \eqc\, 0\leq \xc\leq g_2^{-1} \\ 
K_0+U_1 \frac{2\xc^{s-3}}{s-1} & \eqc\, g_2^{-1}\leq \xc\leq 1 \\
K_0+U_1 \frac{3-s}{(s-1)\xc^2}\left[ 1-g_2^{1-s} \right] & \eqc\, \xc\geq 1  
\end{array} \right. \label{app:implussol2} \eqc
\ee
where we defined $\xc=\gamma_1\tau$, and used the leading term approximation discussed in appendix \ref{sec:appx} for the intermediate regimes. 


\subsection{Large spectral index}

Similarly to the steps in appendix \ref{sec:appx}, we calculate the dependence $\xc(y)$. We begin with the case $s>3$. For $\xc\leq\ggg$ we find
\be
y(\xc\leq\ggg)=\frac{1}{\gamma_1}\int\frac{\td{\xc}}{K_0+U_0(1-g_2^{3-s})} \nonumber \\
=\frac{\xc}{\gamma_1(K_0+U_0(1-g_2^{3-s}))}+\cc_1 \label{ygs3a} \eqd
\ee
As before, we set $\cc_1=0$, since $y(\xc=0)=0$. Inverting Eq. (\ref{ygs3a}) yields
\be
\xc(y\leq\yc_1)=\gamma_1(K_0+U_0(1-g_2^{3-s}))y \label{xcgs3a} \eqd
\ee
Obviously, $\yc_1$ is found from the condition $\xc(y=\yc_1)=\ggg$ yielding 
\be
\yc_1=\ggg(\gamma_1(K_0+U_0(1-g_2^{3-s})))^{-1} \label{xcyc1} \eqd
\ee
For $\ggg\leq\xc\leq 1$ we find
\be
y(\ggg\leq\xc\leq 1)=\frac{1}{\gamma_1}\int\frac{\td{\xc}}{K_0+U_0} \nonumber \\
=\frac{\xc}{\gamma_1(K_0+U_0)}+\cc_2 \label{y1s3a} \eqc
\ee
or the other way around
\be
x(\yc_1\leq y\leq\yc_2)=\gamma_1(K_0+U_0)(y-\cc_2) \label{xc1s3a} \eqd
\ee
Since $\xc$ is supposed to be continuous, we find the constant $\cc_2$ by matching the solutions for $y=\yc_1$ resulting in
\be
\cc_2=\frac{U_0g_2^{2-s}}{\gamma_1(K_0+U_0(1-g_2^{3-s}))(K_0+U_0)} \label{xcc2} \eqd
\ee
We also obtain $\yc_2$ from the condition $\xc(y=\yc_2)=1$
\be
\yc_2=\frac{1}{\gamma_1(K_0+U_0)}\left( \frac{K_0+U_0(1+g_2^{2-s}-g_2^{3-s})}{K_0+U_0(1-g_2^{3-s})} \right) \label{xcyc2} \eqd
\ee 
Defining $B_0=U_0\frac{s-3}{s-1}(1-g_2^{1-s})$ we yield for $\xc\geq 1$
\be
y(\xc\geq 1)=\frac{1}{\gamma_1}\int\frac{\td{\xc}}{K_0+B_0\xc^{-2}} \nonumber \\
= \frac{1}{\gamma_1K_0}\int\td{\xc}\frac{\xc^2}{\frac{B_0}{K_0}+\xc^2} \nonumber \\
= \frac{1}{\gamma_1K_0}\left[ \xc-\sqrt{\frac{B_0}{K_0}}\arctan\left( \sqrt{\frac{K_0}{B_0}}\xc \right)+\cc_3 \right] \label{ys3a} \eqd
\ee
The problem arising is that we cannot find an inverted expression for $\xc$. However, we can obtain approximative results for small and large arguments of the $\arctan$-function. \\
In order to achieve these approximations we define the injection parameter
\be
\alpha_0=\sqrt{\frac{B_0}{K_0}}=\sqrt{\frac{q_0\gamma_1^{3-s}(1-g_2^{1-s})}{K_0(s-1)}} \label{alpha0app} \eqc
\ee
for which Eq. (\ref{ys3a}) becomes
\be
y(\xc\geq 1)=\frac{1}{\gamma_1K_0}\left[ \xc-\alpha_0\arctan\left( \frac{\xc}{\alpha_0} \right)+\cc_3 \right] \label{ys3aA} \eqd
\ee
For $\alpha_0\ll 1$ the argument of the $\arctan$-function is always (much) larger than unity, since $\xc\geq 1$. We, therefore, approximate $\arctan(\xc/\alpha_0)\approx \pi/2$, set $\cc_3'=\cc_3-\pi/2$, and obtain
\be
y(\xc\geq 1,\alpha_0\ll 1)=\frac{1}{\gamma_1K_0}\left[ \xc+\cc_3' \right] \label{ys3a01} \eqd
\ee
This can be easily inverted, yielding the linear solution
\be
\xc(y\geq\yc_2,\alpha_0\ll 1)=\gamma_1K_0(y-\cc_4) \label{xcs3a01} \eqd
\ee
Matching this solution with Eq. (\ref{xc1s3a}) yields
\be
\cc_4=\frac{1}{\gamma_1(K_0+U_0)}\left( \frac{K_0+U_0(1+g_2^{2-s}-g_2^{3-s})}{K_0+U_0(1-g_2^{3-s})} \right) \nonumber \\
-\frac{1}{\gamma_1K_0} \label{xcc4} \eqd
\ee
For $\alpha_0\gg 1$ we have to consider two cases. If $1\leq\xc\leq\alpha_0$, we see that $\xc/\alpha_0<1$. Thus, we can approximate the $\arctan$-function to third order as $\arctan(\xc/\alpha_0)\approx \xc/\alpha_0-\xc^3/3\alpha_0^3$, resulting in
\be
y(1\leq\xc\leq\alpha_0,\alpha_0\gg 1) \nonumber \\
\approx \frac{1}{\gamma_1K_0} \left[ \xc-\alpha_0\left( \frac{\xc}{\alpha_0}-\frac{\xc^3}{3\alpha_0^3} \right)+\cc_3 \right]  \nonumber \\
= \frac{1}{\gamma_1K_0} \left[ \frac{\xc^3}{3\alpha_0^2}+\cc_3 \right] \label{yas3a10} \eqd
\ee
Inverting yields
\be
\xc(\yc_2\leq y\leq\yc_3,\alpha_0\gg 1)=\left[ 3\alpha_0^2\gamma_1K_0(y-\cc_5) \right]^{1/3} \label{xcas3a10} \eqc
\ee
with
\be
\yc_3 = \frac{\alpha_0}{3\gamma_1K_0}+\frac{1}{\gamma_1(K_0+U_0)} \nonumber \\ 
\times \left( \frac{K_0+U_0(1+g_2^{2-s}-g_2^{3-s})}{K_0+U_0(1-g_2^{3-s})} \right) - \frac{1}{3\alpha_0^2\gamma_1K_0} \label{xcyc3} \eqc
\ee
and
\be
\cc_5 = \frac{1}{\gamma_1(K_0+U_0)} \left( \frac{K_0+U_0(1+g_2^{2-s}-g_2^{3-s})}{K_0+U_0(1-g_2^{3-s})} \right) \nonumber \\
- \frac{1}{3\alpha_0^2\gamma_1K_0} \label{xcc5} \eqd
\ee
In the case $\xc\geq\alpha_0$, we can approximate again $\arctan(x/\alpha_0)\approx\pi/2$, yielding with $\cc_3''=\cc_3-\pi/2$
\be
y(\xc\geq\alpha_0,\alpha_0\gg 1)\approx\frac{1}{\gamma_1K_0}\left[ \xc+\cc_3'' \right] \label{ys3a10} \eqc
\ee
or inverted
\be
\xc(y\geq\yc_3,\alpha_0\gg 1) = \gamma_1K_0(y-\cc_6) \label{xcs3a10} \eqc
\ee
with
\be
\cc_6 = \frac{1}{\gamma_1(K_0+U_0)} \left( \frac{K_0+U_0(1+g_2^{2-s}-g_2^{3-s})}{K_0+U_0(1-g_2^{3-s})} \right) \nonumber \\
- \frac{1}{3\alpha_0^2\gamma_1K_0}-\frac{2\alpha_0}{3\gamma_1K_0} \label{xcc6} \eqd
\ee
What one can see here is that the injection parameter controls significantly the cooling behavior of the electrons. For $\alpha_0\ll 1$ the solution is purely linear, while for $\alpha_0\gg 1$ it is non-linear and becomes linear at later times, just as we expected it. \\
Before we proceed with the case $1<s<3$ we list the results of this section once more in a compact form. We also approximate the results for $g_2\gg 1$, which, as one will see, simplifies a lot.
\be
\xc(y\leq\yc_1)&\approx& \gamma_1 (K_0+U_0) y \\
\xc(\yc_1\leq y\leq\yc_2) &=& \gamma_1 (K_0+U_0) (y-\cc_2) \\
\xc(y\geq\yc_2,\alpha_0\ll 1) &=& \gamma_l K_0 (y-\cc_4) \\
\xc(\yc_2\leq y\leq\yc_3,\alpha_0\gg 1) &=& \left[ 3\alpha_0^2\gamma_1K_0(y-\cc_5) \right]^{1/3} \\
\xc(y\geq\yc_3,\alpha_0\gg 1) &=& \gamma_1K_0(y-\cc_6) \eqc
\ee
with
\be
\yc_1 &\approx& \frac{\ggg}{\gamma_1(K_0+U_0)} \\
\yc_2 &\approx& \frac{1}{\gamma_1(K_0+U_0)} \\
\yc_3 &\approx& \frac{1}{\gamma_1(K_0+U_0)}+\frac{\alpha_0^3-1}{3\alpha_0^2\gamma_1K_0} \eqc
\ee
and
\be
\cc_2 &\approx& \frac{U_0g_2^{2-s}}{\gamma_1(K_0+U_0)^2}\rightarrow 0 \\
\cc_4 &\approx& -\frac{U_0}{\gamma_1K_0(K_0+U_0)} \\
\cc_5 &\approx& \frac{(3\alpha_0^2-1)K_0-U_0}{3\alpha_0^2\gamma_1K_0(K_0+U_0)} \\
\cc_6 &\approx& \frac{1}{\gamma_1(K_0+U_0)}-\frac{2\alpha_0^3+1}{3\alpha_0^2\gamma_1K_0} \eqd
\ee
Since in this approximation $\cc_2\rightarrow 0$, $\xc(y\leq\yc_1)=\xc(\yc_1\leq y\leq\yc_2)$, and, thus, one can neglect $\yc_1$.


\subsection{Small spectral index}

We will now derive the explicit form of the implicit time variable $\xc$ for $1<s<3$. \\
The first regime is $\xc\leq\ggg$, yielding
\be
y(\xc\leq\ggg)=\frac{1}{\gamma_1}\int\frac{\td{\xc}}{K_0+U_1(g_2^{3-a}-1)} \nonumber \\
=\frac{\xc}{\gamma_1(K_0+U_1(g_2^{3-s}-1))}+d_1 \label{yg1sa} \eqd
\ee
Since $y(\xc=0)=0$, obviously $d_1=0$, and the inversion becomes
\be
\xc(y\leq\ys_1)=\gamma_1(K_0+U_1(g_2^{3-s}-1))y \label{xcg1sa} \eqc
\ee
with
\be
\ys_1 = \gamma_1(K_0+U_1(g_2^{3-s}-1)) \label{xcys1} \eqd
\ee
The next time step is $\ggg\leq\xc\leq 1$, resulting in
\be
y(\ggg\leq\xc\leq 1) = \frac{1}{\gamma_1}\int\frac{\td{\xc}}{K_0+\frac{2U_1}{s-1}\xc^{s-3}} \nonumber \\
= \frac{s-1}{2\gamma_1U_1}\int\frac{\td{\xc}}{\frac{\beta}{\alpha_0^2}+\xc^{s-3}} \label{y11sa} \eqc
\ee
where we defined $\beta=(3-s)(1-g_2^{1-s})/2$. For $\alpha_0\ll 1$ we see that $\beta/\alpha_0^2\gg 1$ (as long as $s$ is not too close to $1$ or $3$), and with $\xc\leq 1$ we approximate the integral as
\be
y(\ggg\leq\xc\leq 1,\alpha_0\ll 1)\approx \frac{s-1}{2\gamma_1U_1}\int\frac{\alpha_0^2}{\beta}\td{\xc} \nonumber \\
=\frac{(s-1)\alpha_0^2}{2\gamma_1U_1\beta}y+d_2 \label{y11sa01} \eqd
\ee
The inversion is easily performed, yielding
\be
x(\ys_1\leq y\leq\ys_2,\alpha_0\ll 1)=\frac{2\gamma_1U_1\beta}{(s-1)\alpha_0^2}(y-d_2) \label{xc11sa01} \eqc
\ee
where we obtain by matching the solutions
\be
\ys_2 = \frac{(s-1)\alpha_0^2}{2\gamma_1U_1\beta}(1-\ggg)+\frac{\ggg}{\gamma_1(K_0+U_1(g_2^{3-s}-1))} \label{xcys2} \eqc
\ee
and
\be
d_2 = \frac{\ggg}{\gamma_1}\left( \frac{1}{K_0+U_1(g_2^{3-s}-1)}-\frac{(s-1)\alpha_0^2}{2U_1\beta} \right) \label{xcd2} \eqd
\ee
For $\alpha_0\gg 1$ we see that $\beta/\alpha_0^2\ll 1$. As a rough approximation this is also much lower than $\ggg$, and, therefore, we achieve the integral
\be
y(\ggg\leq\xc\leq 1,\alpha_0\gg 1)\approx \frac{s-1}{2\gamma_1U_1}\int\xc^{s-3}\td{\xc} \nonumber \\
=\frac{s-1}{2\gamma_1U_1(4-s)}\xc^{4-s}+d_3 \label{y11sa10} \eqd
\ee
The inverted equation is
\be
\xc(\ys_1\leq y\leq\ys_3,\alpha_0\gg 1) = \left[ \frac{2\gamma_1U_1(4-s)}{s-1}(y-d_3) \right]^{\frac{1}{4-s}} \label{xc11sa10} \eqc
\ee
with
\be
\ys_3 = \frac{s-1}{2\gamma_1U_1(4-s)}(1-g_2^{s-4}) \nonumber \\
+\frac{\ggg}{\gamma_1(K_0+U_1(g_2^{3-s}-1))} \label{xcys3} \eqc
\ee
and
\be
d_3=\frac{\ggg}{\gamma_1(K_0+U_1(g_2^{3-s}-1))}-\frac{(s-1)g_2^{s-4}}{2\gamma_1U_1(4-s)} \label{xcd3} \eqd
\ee
Similarly to the case $\xc\geq 1$ for large spectral indices, we obtain here for the integral in that time regime
\be
y(\xc\geq 1)=\frac{1}{\gamma_1K_0}\int\frac{\xc^2\td{\xc}}{\alpha_0^2+\xc^2} \nonumber \\
=\frac{1}{\gamma_1K_0}\left[ \xc-\alpha_0\arctan\left( \frac{\xc}{\alpha_0} \right)+d_4 \right] \label{y1sa} \eqd
\ee
We will continue with the same approximations as before, yielding for $\alpha_0\ll 1$ $\arctan(\xc/\alpha_0)\approx\pi/2$, and with $d_4'=d_4-\pi/2$ the result
\be
y(\xc\geq 1,\alpha_0\ll 1) \approx \frac{1}{\gamma_1K_0}(\xc+d_4') \label{y1sa01} \eqd
\ee
The inversion is obviously
\be
\xc(y\geq\ys_2,\alpha_0\ll 1)=\gamma_1K_0(y-d_5) \label{xc1sa01} \eqc
\ee
where
\be
d_5=\frac{(s-1)\alpha_0^2}{2\gamma_1U_1\beta}(1-\ggg)+\frac{\ggg}{\gamma_1(K_0+U_1(g_2^{3-s}-1))} \nonumber \\
- \frac{1}{\gamma_1K_0} \label{xcd5} \eqd
\ee
For $\alpha_0\gg 1$ we use for $1\leq\xc\leq\alpha_0$ the approximation $\arctan(\xc/\alpha_0)\approx\xc/\alpha_0-\xc^3/3\alpha_0^3$, yielding
\be
y(1\leq\xc\leq\alpha_0,\alpha_0\gg 1)\approx \frac{1}{\gamma_1K_0}\left[ \frac{\xc^3}{3\alpha_0^2}+d_4'' \right] \label{ya1sa10} \eqd
\ee
Hence,
\be
\xc(\ys_3\leq y\leq\ys_4,\alpha_0\gg 1)=\left[ 3\gamma_1K_0\alpha_0^2(y-d_6) \right]^{1/3} \label{xca1sa10} \eqc
\ee
with
\be
\ys_4 = \frac{\alpha_0^3-1}{3\gamma_1K_0\alpha_0^2}+\frac{(s-1)(1-g_2^{s-4})}{2\gamma_1U_1(4-s)} \nonumber \\
+ \frac{\ggg}{\gamma_1(K_0+U_1(g_2^{3-s}-1))} \label{xcys4} \eqc
\ee
and
\be
d_6 = \frac{(s-1)(1-g_2^{s-4})}{2\gamma_1U_1(4-s)}+\frac{\ggg}{\gamma_1(K_0+U_1(g_2^{3-s}-1))} \nonumber \\
- \frac{1}{3\gamma_1K_0\alpha_0^2} \label{xcd6} \eqd
\ee
The last case is for $\xc\geq\alpha_0$, where we can use the "linear" approximation $\arctan(x/\alpha_0)\approx\pi/2$, again. With $d_4'''=d_4-\pi/2$ we achieve
\be
y(\xc\geq\alpha_0,\alpha_0\gg 1)\approx \frac{1}{\gamma_1K_0}\left[ \xc+d_4''' \right] \label{y1sa10} \eqc
\ee
or inverted
\be
\xc(y\geq\ys_4,\alpha_0\gg 1)=\gamma_1K_0(y-d_7) \label{xc1sa10} \eqc
\ee
where we defined
\be
d_7 = \frac{(s-1)(1-g_2^{s-4})}{2\gamma_1U_1(4-s)}+\frac{\ggg}{\gamma_1(K_0+U_1(g_2^{3-s}-1))} \nonumber \\
- \frac{2\alpha_0^3+1}{3\gamma_1K_0\alpha_0^2} \label{xcd7} \eqd
\ee
As we did for the case of large spectral indices, we sum up our results in a short list, and perform the approximation for $g_2\gg 1$.
\be
\xc(y\leq\ys_1) &\approx& \gamma_1(K_0+U_1g_2^{3-s})y \\
\xc(\ys_1\leq y\leq\ys_2,\alpha_0\ll 1) &=& \frac{2\gamma_1U_1\beta}{(s-1)\alpha_0^2}(y-d_2) \\
\xc(y\geq\ys_2,\alpha_0\ll 1) &=& \gamma_1K_0(y-d_5) \\
\xc(\ys_1\leq y\leq\ys_3,\alpha_0\gg 1) &=& \left[ \frac{2\gamma_1U_1(4-s)}{s-1}(y-d_3) \right]^{\frac{1}{4-s}} \\
\xc(\ys_3\leq y\leq\ys_4,\alpha_0\gg 1) &=& \left[ 3\gamma_1K_0\alpha_0^2(y-d_6) \right]^{1/3} \\
\xc(y\geq\ys_4,\alpha_0\gg 1) &=& \gamma_1K_0(y-d_7) \eqc 
\ee
with
\be
\ys_1 &\approx& \frac{\ggg}{\gamma_1(K_0+U_0g_2^{3-s})} \\
\ys_2 &\approx& \frac{(s-1)\alpha_0^2}{2\gamma_1U_1\beta}+\frac{\ggg}{\gamma_1(K_0+U_0g_2^{3-s})} \\
\ys_3 &\approx& \frac{s-1}{2\gamma_1U_1(4-s)}+\frac{\ggg}{\gamma_1(K_0+U_0g_2^{3-s})} \\
\ys_4 &\approx& \frac{\alpha_0}{3\gamma_1K_0}+\frac{s-1}{2\gamma_1U_1(4-s)}+\frac{\ggg}{\gamma_1(K_0+U_0g_2^{3-s})} \eqc
\ee
and
\be
d_2 &\approx& \frac{\ggg}{\gamma_1(K_0+U_0g_2^{3-s})}-\frac{(s-1)\ggg\alpha_0^2}{2\gamma_1U_1\beta} \\
d_3 &\approx& \frac{\ggg}{\gamma_1(K_0+U_0g_2^{3-s})}-\frac{(s-1)g_2^{s-4}}{2\gamma_1U_1(4-s)} \\
d_5 &\approx& \frac{(s-1)\alpha_0^2}{2\gamma_1U_1\beta}+\frac{\ggg}{\gamma_1(K_0+U_0g_2^{3-s})}-\frac{1}{\gamma_1K_0} \\
d_6 &\approx& \frac{s-1}{2\gamma_1U_1(4-s)}+\frac{\ggg}{\gamma_1(K_0+U_0g_2^{3-s})}-\frac{1}{3\gamma_1K_0\alpha_0^2} \\
d_7 &\approx& \frac{s-1}{2\gamma_1U_1(4-s)}+\frac{\ggg}{\gamma_1(K_0+U_0g_2^{3-s})}-\frac{2\alpha_0}{3\gamma_1K_0} \eqd
\ee


\end{document}